\documentclass[a4paper,fleqn]{cas-sc}

 \usepackage[numbers,sort&compress]{natbib}
 
\usepackage{epsfig}
\usepackage{color}

\usepackage{psfrag}

\newcommand\be{\begin{equation}}
\newcommand\ee{\end{equation}}
\newcommand\q{\quad}
\newcommand\Fr{\text{Fr}}
\newcommand{\dt}{\dot{d}}
\newcommand{\ddt}{\ddot{d}}
\newcommand\C{\mathcal{C}}
\newcommand{\tA}{{\bf{\mathsf{A}}}}

\begin{document}
\let\WriteBookmarks\relax
\def\floatpagepagefraction{1}
\def\textpagefraction{.001}

\shorttitle{Flexibility effects on pitch-heave flutter}

\shortauthors{R. Fernandez-Feria}

\title [mode = title]{Effect of flexibility on the pitch-heave flutter instability of a flexible foil elastically supported on its leading edge}                      
%\tnotemark[1,2]

%\tnotetext[1]{This document is the results of the research
%   project funded by the National Science Foundation.}

%\tnotetext[2]{The second title footnote which is a longer text matter
%   to fill through the whole text width and overflow into
%   another line in the footnotes area of the first page.}

\author[1]{R. Fernandez-Feria}[%type=editor,
                        %auid=000,bioid=1,
                        %prefix=Sir,
                        %role=Researcher,
                        orcid=0000-0001-9873-1933 ]

%\cormark[1]


\ead{ramon.fernandez@uma.es}

%\ead[url]{www.cvr.cc, cvr@sayahna.org}

%\credit{Conceptualization of this study, Methodology, Software}

\affiliation[1]{organization={Fluid Mechanics Group, IMECH.UMA, Universidad de M\'alaga},
    addressline={Dr Ortiz Ramos s/n}, 
    city={M\'alaga},
    % citysep={}, % Uncomment if no comma needed between city and postcode
    postcode={29071}, 
    country={Spain}}
%

\begin{abstract}
An analytical tool is presented to compute the parametric regions  of flutter instabilities  of a two-dimensional flexible foil elastically mounted on  springs and dampers on its leading edge. It is based on a new analytical formulation of the unsteady fluid-estructure interaction valid for  small-amplitude oscillations and deformations of the foil immersed in an inviscid fluid current. The formulation extends a previous analysis 
by including the effects of gravity and a second flexural mode, increasing its validity range to much smaller rigidities. The analytical results are validated with available numerical results, capturing almost exactly the first two natural flexural modes of a clamped foil down to values of the stiffness parameter $S$ of order $10^{-1}$,  but not covering flapping-flag instabilities occurring at smaller values of $S$. When only passive heave, or only passive pitch, is allowed at the leading edge, the rigid foil is stable, as is well known, existing an upper stiffness bound for the flexural instabilities, characterized here as the linear, or torsional, spring constant decreases from infinity. For small enough values of the spring constant, the flexural instability mode  becomes coupled with the spring instability mode, increasing the growth rate substantially. These coupled spring (linear or torsional) and flexural  instability modes occur below a threshold value of $S$ and above a threshold value of $R$, both depending on the corresponding spring constant.  Coupled pitch-heave flutter instabilities of a rigid  are possible in a region below a curve of the parametric plane of the two springs constants that depends on $R$, which shrinks  to zero as $R$ decreases.  For a flexible foil, the flexural unstable modes become coupled with the springs unstable mode as $S$ decreases from infinity, enlarging the mass ratio range for flutter instability and increasing its growth rate, the more so the smaller the springs constants. All these parametric regions for  flutter instabilities are easily characterized with the present analytical tool, providing the corresponding frequency and critical flutter fluid velocity for the different possible configurations. Thus, the present results can be useful, for instance, as a guide in the design and sizing of future  turbines based on flexible oscillating foils.

%\noindent\texttt{\textbackslash begin{abstract}} \dots 
%\texttt{\textbackslash end{abstract}} and
%\verb+\begin{keyword}+ \verb+...+ \verb+\end{keyword}+ 
%which
%contain the abstract and keywords respectively. 
%
%\noindent Each keyword shall be separated by a \verb+\sep+ command.
%\end{abstract}

% \includegraphics{figs/grabs.pdf}

%% Research highlights
%\item Developed an analytical tool to predict flutter instability regions of a flexible foil mounted on leading-edge springs and dampers.
%\item Extended previous model by including gravity and a second flexural mode, improving accuracy for lower stiffness values.
%\item Identified coupled springs-flexural instability regions and critical flutter conditions, providing guidance for flexible foil turbine design.

% Keywords

\maketitle

\section{Introduction} \label{sec_intro}

Characterizing the onset of flutter instability of a flexible plate in uniform flow is both a classical theoretical problem in fluid-structure interaction and a relevant problem of interest in many engineering fields. 
Flutter instabilities, arising above a critical free stream fluid velocity from the dynamic interaction between aerodynamic forces, structural elasticity, and inertia, can lead to self-excited oscillations that grow in amplitude and potentially result in structural failure.
While flutter of flexible foils or plates has traditionally been treated as a limiting instability in aeroelastic design, specially in aeronautics  \cite{fung69,bisas75} and some manufacturing processes \citep{watsu02,leele06}, it also serves to understand some physiological phenomena \citep{huang95,paido16}, and recent research has demonstrated its potential as an effective mechanism for flow-induced energy harvesting \citep{tanpa09,doami11,dunst11}. In particular, two-dimensional foils provide a canonical and well-controlled configuration for investigating the fundamental physics of flutter-driven energy extraction through fully passive flapping foil turbines \cite{boudu18,duade19,boupi20,goypo23,tobaz25}.

Owing to their mathematical tractability and physical clarity, two-dimensional (2-D) foil models have long been used as the basis for analytical aeroelastic theories, numerical simulations and laboratory experiments enabling systematic investigation of flutter onset and post-critical behavior \cite{theod35,kordo76,huang95,guopa00,sheva05,argma05,eloso07,tanpa07,elola08,micll08,albsh08,alben08a,chawa12}. Within the vast literature on flutter instabilities in 2-D foils,  this work is focused on analytical theories that take into account foil flexibility to characterize its effect on pitch-heave coupled flutter of a flexible plate. 

Classical analytical descriptions of foil flutter are rooted in linear potential-flow theory. Theodorsen's theory, in particular, provides a closed-form expression for the lift and moment acting on an oscillating airfoil undergoing harmonic pitching and plunging motions \citep{theod35}. When coupled with linear structural models, this formulation yields an eigenvalue problem whose solution predicts flutter speed, frequency, and mode coupling. Despite its simplifying assumptions, this approach remains a cornerstone of flutter analysis and continues to inform the design of flutter-based energy harvesting systems, allowing parametric studies of mass ratio, stiffness, and energy extraction mechanisms.
More recently, data-driven aerodynamic models are also being commonly used for aeroelastic simulations and flutter stability analyses \citep{lelho23}. They are based on high fidelity CFD simulations and usually provide  precise models valid even for large amplitudes of deformations and oscillations. However, they require costly numerical simulations for their development and are only valid for  a specific fluid-structure configuration, without reliable information about their validity when the configuration changes. Models based on linear potential theory, although valid only for small amplitudes, adapt easily to different configurations and it is always possible to quantify their validity range when additional simplifications are made because the theory provides a precise physical meaning of each contribution to the model. In any case, descriptions based on linear potential flow theory always provide a first direct analytical estimate of the onset and the physics of flutter instability.

Theodorsen's flutter theory is for a rigid foil with passive pitch and heave. To account for the effect of the flexibility of the foil, one needs to solve the fluid-structure interaction (FSI) equations coupling the linearized inviscid flow equations with the equation for the plate dynamics for appropriate boundary conditions. The resulting complex eigenvalue problem has been formulated and solved numerically in a number of works, mostly using Galerkin decomposition methods, but limited to cases with a fixed leading edge,  specially for a clamped-free plate (clamped at its leading edge and free at its  trailing edge) \citep[e.g.,][]{eloso07,alben08a} or a pinned-free plate \citep[e.g.,][]{chawa12}, though other boundary conditions  have also been considered \citep[e.g.,][]{guopa00}. To obtain closed analytical descriptions one needs further simplifying  approximations for the FSI. Still for a plate with a fixed leading edge, Kornecki et al. \cite{kordo76}  used a quasi-steady, non-circulatory  aerodynamic theory and solved the resulting simplified eigenvalue problem using two modes only to obtained an approximation of the critical parameters for the flutter instability of a clamped-free plate. A simpler approach was considered by Shelley et al. \cite{sheva05} for a heavy foil in water, assuming a superposition of sinusoidal traveling waves for the foil and the inviscid fluid and considering a linear stability analysis, finding a simple expression for the flutter velocity. For the same problem with a clamped-free foil, Argentina and Mahadevan \cite{argma05} included the circulatory part of the inviscid flow loads and considered the limit of vanishing mass ratio $R$ and large non-dimensional stiffness $S$ of the foil (the non-dimensional parameters $R$ and $S$ will  be defined below in Eq, \eqref{RandS}), finding that this is also equivalent to a quasi-steady approximation with Theodorsen function equal to unity, and providing  a threshold for stability.

A different analytical approach, which also incorporated the classical coupled-mode flutter by allowing for passive heave and pitch at the leading edge of the foil, 
was considered in \cite{ferna22}.  This approach  made use of closed-form analytical expressions for the unsteady lift, moment and the first flexural moment exerted by the inviscid flow on an flexible pitching and heaving foil described by a general fourth-order polynomial \cite{alafe20,feral21}. The closed-forms of the  force and moments were coupled with  the first three moments of the Euler-Bernoulli (E-B) beam equation to yield a simple FSI analytical formulation, resulting in an eigenvalue problem which provided a simple and efficient  tool to characterize the onset of flutter instability in terms of the different non-dimensional parameters such as mass ratio, center-of-mass location, the stiffnesses of the torsional and linear springs supporting the foil, and the stiffness of the flexible foil. These analytical results were validated with experimental data of flutter-based energy harvesting systems using elastically supported rigid foils, as well as with numerical results for a clamped-free  flexible foil when its stiffness was  sufficiently high. But the analytical formulation failed when the stiffness parameter $S$ dropped below about ten.

To extend this analytical approach to smaller stiffnesses of the foil, un thus uncovering new instability modes,  one has to derive new analytical expression for higher flexural moments of the flexible foil and consider higher moments of the E-B equation. This may be a cumbersome task, with probably no clear advantages with the classical numerical methods to solve the FSI problem  based on, for instance,  Galerkin decomposition methods. However, using just one further (second) flexural moment, the analytical FSI formulation based on the first four moments of the E-B equation provides a very accurate approximation to the first two natural bending modes of the flexible plate \citep{ferna25}, and, as it will shown here, considerably extends the validity range of the flutter stability analysis down to values of the  stiffness parameter $S$ of $O(10^{-1})$. With this extension, we shall see that the new analytical model uncovers additional unstable modes which were not provided by the formulation in \cite{ferna22}. 
Of course, there are still many flutter instabilities for $S \lesssim 10^{-1}$ that are relevant in some situations (the so-called flapping-flag instabilities \cite[e.g.,][]{alben08a}). But this range is more than enough in many engineering applications, especially in the design of energy extraction systems based on pitch-heave coupled flutter of {\it{flexible}} foils.

The main objective of the present work is therefore to use an reformed version of  the analytical FSI formulation derived in \cite{ferna25}  to analyze the flutter instabilities of an elastically mounted flexible foil through spring and dampers on its leading edge. This formulation, though still limited to not too small stiffnesses, and thus not accounting for many flapping-flag flutter instabilities appearing in very flexible foils,  provides a  convenient analytical tool for quick estimation of flutter instabilities in many situations  of physical and engineering  interest, specially for flexible foils with passive pitch and/or heave. The problem is formulated in \S \ref{sec_for}, were the two-dimensional foil is assumed horizontal and the effect of gravity, not considered in \cite{ferna22}, is also included, obtaining the foil equilibrium location and shape in a fluid at rest and  immersed in a fluid current. The flutter stability analysis and its validation with previous numerical results for a clamped foil are considered in \S \ref{sec_flutter}. Flutter instability results for a flexible foil coupled with passive pitch-only, passive heave-only, and coupled pitch-heave are presented and discussed in \S \ref{sec_results}, where the results are also validated against previous numerical results for a pinned-free  plate. Finally, key conclusions are summarized in \S \ref{sec_conclu}.

\begin{figure}
  \begin{psfrags}
    \psfrag{x}[l][l][1]{$x$}
      \psfrag{h}[l][l][1]{$h(t)$}
        \psfrag{a}[l][l][1]{$\alpha(t)$}
    \psfrag{e}[l][l][1]{$\varepsilon$}
    \psfrag{c}[l][l][1]{$c$}
    \psfrag{kh}[l][l][1]{$k_h, b_h$}
    \psfrag{ka}[l][l][1]{$k_a, b_a$}
    \psfrag{z}[l][l][1]{$z_s(x,t)$}
 \psfrag{U}[l][l][1]{$U$}
   \centerline{\epsfig{file=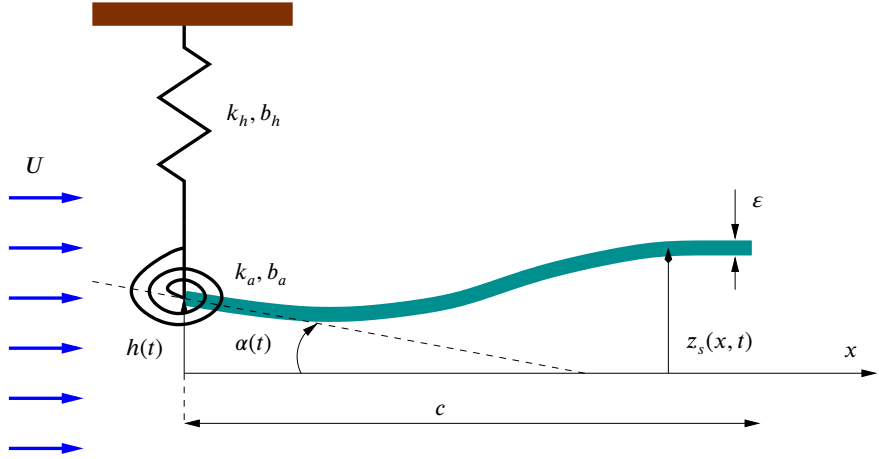,width=0.7\linewidth}}
  \end{psfrags}
  \caption{Schematic of the problem.}
  \label{fig_sketch}
\end{figure}

\section{Formulation of the problem}
\label{sec_for}

Taking into account the effect of gravity, with acceleration $g$ in the $-z$ direction, the E-B beam equation for a 2-D thin and inextensible foil  immersed in an inviscid flow can be written, in the linear approximation for small-amplitude of its motion and deformation around the plane $z=0$, as  \citep[e.g.,][]{doyle01,feral21}:
\be \rho_s \varepsilon \frac{\partial^2 z_s}{\partial t^2} + \frac{\partial^2 }{\partial x^2} \left( EI \frac{\partial^2 z_s}{\partial x^2} \right) = \Delta p + F_{p} \delta(x-x_p) -M_{p} \delta'(x-x_p) - \rho_s \varepsilon g\,, \label{E-B}
\ee
where $t$ is the time, $x$ the coordinate along the unperturbed foil of chord length $c$,  with $-c/2 \leq x \leq c/2$, $\varepsilon$ is the foil  thickness and $z_s(x,t)$ the displacement of its centerline, both satisfying  $\varepsilon \ll c$ and  $|z_s(x,t)| \ll c$; $\rho_s$ and $E I = E \varepsilon^3/12$ are the solid density and structural bending rigidity per unit span of the foil, respectively, and $\Delta p = p^- - p^+$ is the fluid pressure difference between the lower and upper sides of the foil, which is elastically supported on a given pivot axis located at $x=x_p$ through linear and torsional springs and dampers. Thus, the point force and moment (per unit span) at the pivot axis, $F_p$ and $M_p$, are given by
\be F_p= - k_h h - b_h \dot{h} \,, \q M_p = k_a \alpha + b_a \dot{\alpha} \,, \label{spring-damper}
\ee
where $h(t)$ and $\alpha(t)$ are the passive vertical displacement (heave)  and  rotation angle (pitch) of the foil at the pivot axis, respectively, $\dot{h}$ and $\dot{\alpha}$ are their temporal derivatives, whereas $k_h$, $b_h$, $k_a$ and $b_a$ are the linear and torsional springs  and dampers constants. Note that  the moment is taken positive when counterclockwise, while $\alpha$ is positive when clockwise, as it is commonly used in aerodynamics (see Fig. \ref{fig_sketch}). In the present analysis the pivot axis will be located at, or very close to, the leading edge; i.e.  $x_p=-c/2$.

\subsection{Non-dimensional \textit{moment} equations}

From now on, non-dimensional quantities will be used, unless otherwise specified, using the half-chord length $c/2$, the free streaming velocity $U$ in the $x-$direction and the fluid density $\rho$ as scaling parameters. For the sake of notation simplicity, the same symbols will be used for the non-dimensional $x$ (varying now between $-1$ and $1$), time $t$, displacement $z_s$,  the heave $h$ (the pitch $\alpha$ is already dimensionless), and the springs and dampers constants in Eq. \eqref{spring-damper} when these expressions are non-dimensionalized.

In order to be able to compute the fluid force and moments analytically for small-amplitude oscillations and deformations of the foil, capturing up to the second natural bending mode of the system, a fifth-order approximation of the foil motion is assumed \citep{ferna25}:
\[ z_s(x,t)= h(t) - \alpha(t) (x+1) + d_1(t) [ 24(x+1)^2-8(x+1)^3+(x+1)^4]  \]
\be \q\q\q\q\   + d_2(t)[160(x+1)^2 -40(x+1)^3+(x+1)^5] \,, \q\q\q  -1 \leq x \leq 1 \,, \label{zs_def}
\ee 
where $d_1(t)$ and $d_2(t)$ are  the first two non-dimensional flexural deformation modes, which, like the heave and pitch $h(t)$ and $\alpha(t)$, are all \em passive \rm (i.e., unknowns) in the present work, generated by the fluid motion and the effect of gravity and taking into account the inertia of the foil. This expression satisfies the pitch-heave boundary condition at the leading edge, $z_s=h(t)$, $\partial z_s/\partial x = -\alpha(t)$ at $x=-1$, and free trailing edge,  $\partial^2 z_s/\partial x2 =  \partial^3 z_s/\partial x^3=0$ at $x=1$.  

Using \eqref{zs_def} in Eq. \eqref{E-B}, the first four \em moments \rm of the E-B equation, i.e., multiplying \eqref{E-B} in non-dimensional form by $1$, $x+1$, $(x+1)^2$ and $(x+1)^3$, and integrating between $x=-1$ and $x=1$, are (assuming  that $\rho_s$, $\varepsilon$ and $E$ are constant along the foil's chord length):
\be R \left( \ddot{h} + \frac{1}{2 \Fr^2} - \ddot{\alpha} + \frac{96}{5} \ddot{d}_1 + \frac{416}{3} \ddot{d}_2 \right) =C_L+C_{L_p} \,, \label{mom1a}
\ee
\be R \left( \frac{1}{2} \ddot{h} + \frac{1}{4 \Fr^2} - \frac{2}{3} \ddot{\alpha} + \frac{208}{15} \ddot{d}_1 + \frac{704}{7} \ddot{d}_2 \right) =C_M+C_{M_p} \,, \label{mom2a}
\ee
\be R \left( \frac{4}{3} \ddot{h} + \frac{2}{3 \Fr^2} - 2 \ddot{\alpha} + \frac{4544}{105} \ddot{d}_1 + \frac{944}{3} \ddot{d}_2 \right) + S \left( \frac{32}{3} d_1 + 80 d_2 \right) = C_{F_1}  \,, \label{mom3a}
\ee
\be R \left( 2 \ddot{h} + \frac{1}{\Fr^2} - \frac{16}{5} \ddot{\alpha} + \frac{496}{7} \ddot{d}_1 + \frac{32 512}{63} \ddot{d}_2 \right) + S \left( 16 d_1 + 128 d_2 \right) = C_{F_2}  \,, \label{mom4a}
\ee
\noindent where dots are used for the derivatives with respect to the non-dimensional time $t$. In these equations,   the non-dimensional parameters 
\be R = \frac{4 \rho_s \varepsilon}{\rho c} \q\q \text{and} \q \q S= \frac{4 E \varepsilon^3}{\rho U^2 c^3} \label{RandS}
\ee
are the mass ratio (or inertia parameter) and the bending stiffness parameter, respectively, and the following fluid force and moment coefficients have been defined:
\be C_L = 
\int_{-1}^1 (\Delta P) d x \,, \q\q \text{with} \q\q \Delta P = \frac{\Delta p}{\rho U^2} \,, \label{defCL}
\ee
\be  C_M =  
 \frac{1}{2}  \int_{-1}^1 (x+1) (\Delta P) d x  \,, \q C_{F_1} = 
\int_{-1}^1 (x+1)^2 (\Delta P) d x 
\,, \q C_{F_2} = 
\int_{-1}^1 (x+1)^3 (\Delta P) d x  \,, 
\label{defCF2}
\ee
\noindent together with the point force and moment coefficients associated to the linear and torsional springs and dampers at the leading edge \eqref{spring-damper},
\be C_{L_p} = \frac{F_{p}}{\frac{1}{2} \rho U^2 c} = - k_h h - b_h \dot{h}  \,, \q C_{M_p} =  \frac{M_p}{\frac{1}{2} \rho U^2 c^2} = k_a \alpha + b_a \dot{\alpha}  \,. \label{CLpCMp}
\ee
The analytical expressions of the  coefficients \eqref{defCL}-\eqref{defCF2} for a harmonic motion of the foil  are given in Appendix \ref{app_coeffi}. The non-dimensional spring and damper constants $k_h$, $b_h$, $k_a$ and $b_a$ are made dimensionless with $\rho U^2$, $\rho U c/2$, $\rho U^2 c^2/2$ and $\rho U c^3/4$, respectively (remember that all  the forces and moments are defined per unit span).

The effect of gravity enters into the equations \eqref{mom1a}-\eqref{mom4a} through the Froude number
\be \Fr = \frac{U}{\sqrt{c g}} \,, \label{Froude}
\ee
which  has been placed on the left side of the equations to appreciate that its effect is equivalent to a dimensionless  effective heave given by
\be h_{eff} = h(t) + \frac{t^2}{4 \Fr^2} \,, \q\q  \text{with acceleration} \q\q \ddot{h}_{eff} = \ddot{h} + \frac{1}{2 \Fr^2} \,. \label{heff}
\ee

\subsection{Static equilibrium in a fluid at rest}
When the elastically supported foil is immersed in a fluid at rest, one can find its equilibrium location and deformation, characterized by the non-dimensional constants $h_e$, $\alpha_e$, $d_{1e}$ and $d_{2e}$, by using the (non-dimensional) static pressure distribution 
\be (\Delta P)_e = \frac{(p^- - p^+)_e}{\rho U^2} = \frac{\rho g \varepsilon}{\rho U^2} = \frac{\varepsilon}{c} \frac{1}{\Fr^2} \label{pest}
\ee
to compute the fluid force and moments (buoyancy effects): 
\be C_{L_e}=  \frac{2 \varepsilon}{c} \frac{1}{\Fr^2} \,, \q  C_{M_e}=  \frac{\varepsilon}{c} \frac{1}{\Fr^2} \,, \q 
C_{F_{1e}}=  \frac{8 \varepsilon}{3 c} \frac{1}{\Fr^2} \,, \q  C_{F_{2e}}=  \frac{4 \varepsilon}{c} \frac{1}{\Fr^2} \,. \label{CFe}
\ee
Note that, although now $U=0$, the pressure, forces and moments are still non-dimensionalized using $U$ to maintain the same notation, but, of course, the final static equilibrium results are independent of $U$ (see below).

\begin{figure}
\centerline{\epsfig{file=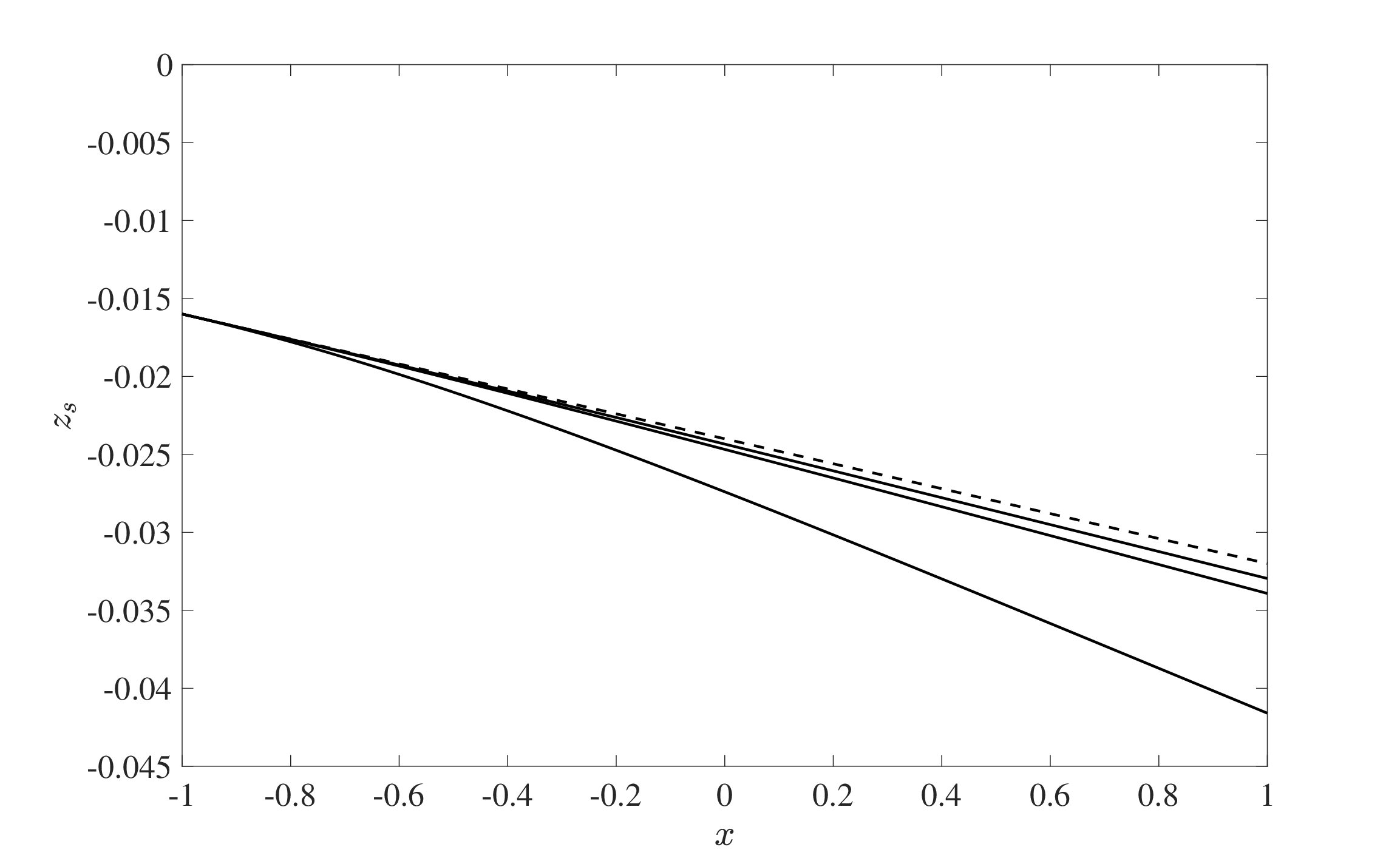,width=.75\linewidth}}
  \caption{Equilibrium foil location and shape $z_{s_e}(x)$ for $G=0.16$ ($R=0.1$, $\Fr=0.5$ and $\rho/\rho_s=0.2$), $k_h=k_a=10$ and, from bottom to top, $S=100, 500,$ and $1000$. The dashed straight line corresponds to a rigid foil ($S\to \infty$).}
\label{fig_equi}
\end{figure}

Substituting these expressions into Eqs.  \eqref{mom1a}-\eqref{mom4a}, together with \eqref{CLpCMp}, for the constant values $h=h_e$, $\alpha=\alpha_e$, $d_1=d_{1e}$ and $d_2=d_{2e}$, one obtains the following equations:
\be  \frac{R}{2 \Fr^2} = \frac{2 \varepsilon}{c} \frac{1}{\Fr^2} - k_h h_e\,, \label{mom1ae}
\ee
\be  \frac{R}{4 \Fr^2}  = \frac{\varepsilon}{c} \frac{1}{\Fr^2} + k_a \alpha_e  \,, \label{mom2ae}
\ee
\be  \frac{2 R}{3 \Fr^2} + S \left( \frac{32}{3} d_{1e} + 80 d_{2e} \right) = \frac{8 \varepsilon}{3 c} \frac{1}{\Fr^2} \,, \label{mom3ae}
\ee
\be \frac{R}{\Fr^2} + S \left( 16 d_{1e} + 128 d_{2e} \right) = \frac{4 \varepsilon}{c} \frac{1}{\Fr^2} \,, \label{mom4ae}
\ee
so that the static equilibrium of the foil is characterized by
\be h_e = - \frac{G}{k_h} \,, \q \alpha_e= \frac{G}{ 2 k_a} \,, \q d_{1e}= - \frac{G}{8 S} \,, \q d_{2e} = 0 \,, \label{equipos}
\ee
where a new non-dimensional parameter relating gravity and inertia has been defined:
\be G = \frac{R}{2 \Fr^2} \left( 1 - \frac{\rho}{\rho_s} \right) = \frac{2 \varepsilon g (\rho_s - \rho)}{\rho U^2} \,. \label{defG}
\ee
It is worth noticing that $d_{2e}=0$, so that the equilibrium deformation is just characterized by the first flexural deflection mode $d_1$, implying that this equilibrium solution is exact within the present linearized theory, i.e.,  it is not affected by  the  approximation \eqref{zs_def} for $z_s$ since  a higher-order approximation will yield the same equilibrium result. Also, as mentioned above,  the equilibrium values \eqref{equipos} for a fluid at rest are obviously independent of the fluid velocity $U$, for, although the parameter $G$ contains $U^2$ in the denominator, so do have it the non-dimensional quantities $k_h$, $k_a$ and $S$. In fact, using the definitions of the non-dimensional parameters $G$, $k_h$, $k_a$ and $S$, the equilibrium values \eqref{equipos} can be written as
\be h_e = - \frac{2 \varepsilon g (\rho_s - \rho)}{\tilde{k}_h} \,, \q \alpha_e= \frac{c^2 \varepsilon g (\rho_s - \rho)}{ 2 \tilde{k}_a} \,, \q d_{1e}= - \frac{c^3 g (\rho_s - \rho)}{16 \varepsilon^2 E} \,, \q d_{2e} = 0 \,, \label{equiposd}
\ee
where a tilde is used to denote the dimensioned counterparts of $k_h$ and $k_a$. 
Figure \ref{fig_equi} shows the equilibrium location and shape of the foil in a fluid at rest,
 \be z_{s_e}(x) = h_e - \alpha_e (x+1) + d_{1e} [ 24(x+1)^2-8(x+1)^3+(x+1)^4] \,, \label{zsequi}
 \ee 
 for a typical 2-D foil in water, $\rho/\rho_s=0.2$, $R=0.1$, $\Fr=0.5$ (i.e., $G=0.16$), with $k_h=k_a=10$ and several values of the bending stiffness $S$.

\subsection{Static equilibrium in a fluid current}
\label{sec_equiflu}
If the foil is immersed in a fluid current with velocity $U$ in the $x$ direction, it may also reach a new stationary equilibrium position and shape if the new fluid forces and moments balance  gravity and the springs forces and moments for some constant values $h=h^*_{e}$, $\alpha=\alpha^*_{e}$, $d_1=d^*_{1e}$ and $d_2=d^*_{2e}$. 

To obtain these new equilibrium parameters one needs the fluid force and moment coefficients for a stationary foil in terms of constant values of $h$, $\alpha$, $d_1$ and $d_2$ in $z_s(x)$. These fluid flow loads can be derived from the expressions reported in Appendix \ref{app_coeffi}, which, though given for a harmonic motion of the foil, yield the case of a stationary foil with $z_s=z_s(x)$ by just setting to zero all the temporal derivatives and $k =0$ in Theodorsen's function (see Appendix \ref{app_coeffi_e}). The resulting force and moment coefficients mainly come from their circulatory terms associated to $\Gamma_0$:
\be C^*_{L_e}= \pi \left(  2 \alpha - 59 d_1  - \frac{1755}{4} d_2 \right) \,, \label{CLa}
\q\q C^*_{M_e} \simeq \frac{\pi}{4} \left(  2 \alpha - 59 d_1  - \frac{1755}{4} d_2 \right) \,, 
\ee
\be C^*_{F_{1e}} \simeq \frac{\pi}{2} \left(  2 \alpha - 59 d_1  - \frac{1755}{4} d_2 \right) \,, 
\q\q  C^*_{F_{2_e}} \simeq \frac{5 \pi}{8} \left(  2 \alpha - 59 d_1  - \frac{1755}{4} d_2 \right) \,. \label{CF2a}
\ee
Adding to these coefficients the buoyancy force coefficients \eqref{CFe} and substituting  into  \eqref{mom1a}-\eqref{mom4a}, together with \eqref{CLpCMp} for the lineal and torsional springs, one obtains a set of four linear equations whose solution yields the following new equilibrium values in a fluid current with velocity U:
\be h_e^* = \left(\frac{G}{k_h}\right) \, \frac{ -1 + \pi \left( \frac{1}{2 \, k_a} + \frac{1165}{256 \, S} \right)}{ 1 + \pi \left( \frac{1}{2 \, k_a} + \frac{723}{256 \, S} \right)} \,, \label{hea}
\q\q \alpha_e^* = \left(\frac{G}{2 k_a}\right) \, \frac{ 1 - \pi  \frac{221}{256 \, S} }{ 1 + \pi \left( \frac{1}{2 \, k_a} + \frac{723}{256 \, S} \right)} \,,
\ee
\be d_{1e}^* = \left(\frac{G}{8 S}\right) \, \frac{ -1 + \pi \left( \frac{13}{16 \, k_a} + \frac{1755}{256 \, S} \right)}{ 1 + \pi \left( \frac{1}{2 \, k_a} + \frac{723}{256 \, S} \right)} \,, \label{d1ea}
\q\q d_{2e}^* = - \pi \, \left(\frac{G}{2 S}\right) \, \frac{  \frac{1}{32 \, k_a} +  \frac{59}{256 \, S} }{ 1 + \pi \left( \frac{1}{2 \, k_a} + \frac{723}{256 \, S} \right)} \,. 
\ee
Comparing these expressions with \eqref{equipos} for a fluid at rest it is clear that the corrections due to the fluid motion are the terms multiplied by $\pi$, which in the case of  $d_{2e}^*$ are the only ones.

\begin{figure}
\centerline{\epsfig{file=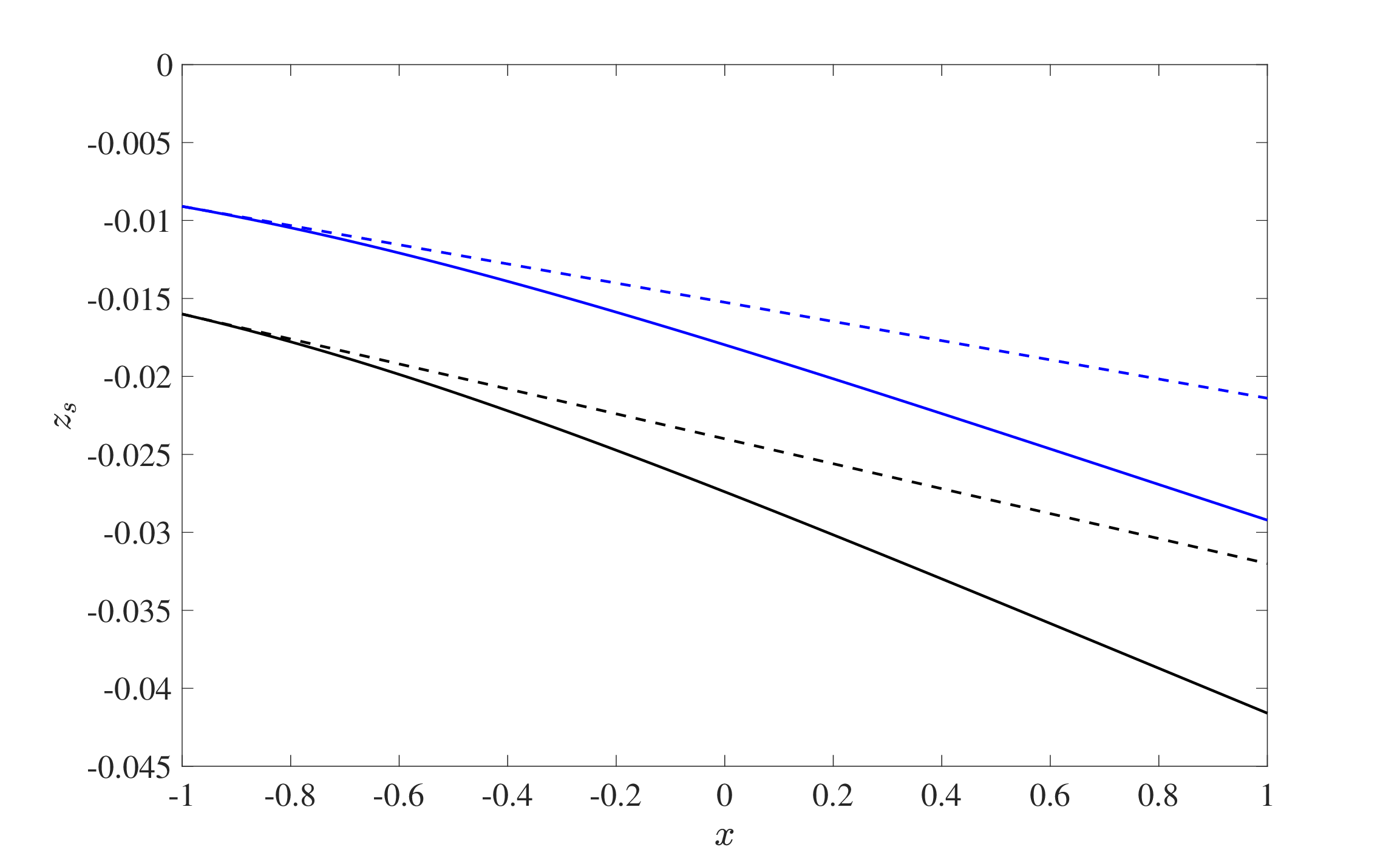,width=.75\linewidth}}
  \caption{Comparison between  $z_{s_e}(x)$ given by \eqref{zsequi} for a foil in a fluid at rest (black lines) with $z_{s_e}^*(x)$ given by \eqref{zsequia} for a foil in a fluid current (blue lines), for the case with $S=100$ in Fig. \ref{fig_equi}. Dashed straight lines correspond to rigid foils.}
\label{fig_equi2}
\end{figure}

Actually, according to the expressions given in Appendix \ref{app_coeffi}, there are additional stationary contributions in the non-circulatory parts of $C_M$, $C_{F_1}$ and $C_{F_2}$ (but not in $C_L$) associated to constant values of the flexural deformation modes $d_1$ and $d_2$, which are included in the coefficients for a stationary foil written in Eqs. \eqref{heaa}-\eqref{d2eaa} of Appendix \ref{app_coeffi_e}. These terms just add very small corrections to the equilibrium values \eqref{hea}-\eqref{d1ea}. Since the corresponding expressions are more involved they are reported in Appendix \ref{app_coeffi_e}. In  the computations of the new equilibrium foil location and shape,
\be z_{s_e}^*(x) = h_e^* - \alpha_e^* (x+1) + d_{1e}^* [ 24(x+1)^2-8(x+1)^3+(x+1)^4]  + d_{2e}^*[160(x+1)^2 -40(x+1)^3+(x+1)^5]  \,, \label{zsequia}
 \ee 
we use these more accurate expressions of $h^*_{e}$, $\alpha^*_{e}$, $d^*_{1e}$ and $d^*_{2e}$ reported in Appendix \ref{app_coeffi_e} (the differences with \eqref{hea}-\eqref{d1ea} are very small indeed).
Figure \ref{fig_equi2} compares the equilibrium location and shape of a foil  immersed in a fluid current, given by \eqref{zsequia}, with that corresponding to the same  foil in the same fluid at rest, given by \eqref{zsequi},  for the case with $S=100$ plotted in Fig. \ref{fig_equi}. It is observed that the fluid motion reduces both the heave and pitch equilibrium displacements, and also its flexural deformation, an obvious consequence of the larger fluid forces and moments which act, like the buoyancy forces,  in the oposite direction of gravity. 

It is also worth commenting that the equilibrium solution \eqref{hea}-\eqref{d1ea} (or \eqref{heaa}-\eqref{d2eaa}) for $h^*_{e}$, $\alpha^*_{e}$, $d^*_{1e}$ and $d^*_{2e}$ does exist for any set of values of the non-dimensional parameters (i.e., the denominators in these expressions never vanish). This means that no divergence instabilities are possible for this flexible foil configuration for any fluid velocity, which was to be expected because the pivot axis is at the leading edge while the center of mass is at the middle of the foil \cite[e.g., Ref. ][chap. 3]{fung69}. However, even if this static equilibrium solution exists for any set of values of the non-dimensional parameters, the foil may undergo flutter instabilities in some parametric ranges, whose characterization is the subject of the present work, to be analyzed next.

\section{Flutter stability analysis}
\label{sec_flutter} 
To find out the possible flutter instabilities we proceed similarly to  \cite{ferna22}, but now taking into account the effect of gravity and considering an additional flexural deformation mode, which modify substantially the parametric ranges of the instabilities. This method was pioneered by \cite{theod35}, who first obtained analytically the fluid force and moment for a rigid foil undergoing harmonic pitching and heavy motions to solve the equilibrium force and moment equations on the airfoil and derive the parametric ranges for flutter instabilities, particularly the flutter velocity in terms of the different non-dimensional parameters. Here we use the analytical expressions for the fluid force, moment and flexural moments on a flexible foil with a general motion and deformation described by \eqref{zs_def},  obtained in \cite{ferna25} and summarized in Appendix \ref{app_coeffi}, to analyze the stability of the foil using  Eqs. \eqref{mom1a}-\eqref{mom4a}. This set of equations, which comes from the first four moments of the E-B equation, generates an analytical eigenvalue problem that provides the flutter characteristics of a flexible foil supported elastically on its leading edge.

Assuming that the foil undergoes small-amplitude perturbations about its equilibrium shape \eqref{zsequia}, to analyze the stability of this equilibrium solution one looks for solutions with heave,  pitch and deformation modes of the form
\be  h(t) = h_e^*+ \Re \left[ H e^{i \gamma t} \right] \,, \q \alpha(t) = \alpha_e^* + \Re \left[ A e^{i \gamma t} \right] \,, \q  d_1(t) = d_{1e}^* +\Re \left[ D_1 e^{i\gamma t} \right] \,, \q  d_2(t) = d_{2e}^* +\Re \left[ D_2 e^{i\gamma t} \right]  \,,\label{halphad} \ee
where $H$, $A$, $D_1$, $D_2$ and $\gamma$ are complex quantities, with
\be  \gamma = k + i \sigma \,, \q\q  k=\frac{\omega c}{2U} \,, \label{defk} \ee
being $k$ the reduced frequency and minus $\sigma$ the non-dimensional growth rate.The absolute values of $H$, $A$, $D_1$ and $D_2$ are the corresponding amplitudes of heave, pitch and deformations perturbations,  whereas their angles define their respective phase shifts. Substituting these expressions into Eqs. \eqref{mom1a}-\eqref{mom4a} together with the fluid force and moments coefficients $C_L$, $C_M$, $C_{F_1}$ and $C_{F_2}$, containing the components corresponding to the buoyancy forces \eqref{CFe}, the stationary part \eqref{clea}-\eqref{cfea} corresponding to the foil's equilibrium  shape, and the unsteady components \eqref{CL0}-\eqref{G0t} corresponding to its harmonic motion (using Theodorsen's function evaluated at $\gamma$,  $\C(\gamma)$), and, finally, the expressions \eqref{CLpCMp} for the linear and torsional springs and dampers, the terms containing the constant part in \eqref{halphad}  cancel out and one obtains a linear system of four algebraic equations for the complex amplitudes $H$, $A$, $D_1$ and $D_2$:
\be \tA(\gamma) \cdot {\bf{X}} = {\bf{0}} \,,  \q\q  \text{with} \q\q  \tA =   \left( \begin{array}{cccc}
                                                             A_{hh} & A_{ha} & A_{ha} & A_{h2} \\
                                                            A_{ah} & A_{aa} & A_{a1} & A_{a2} \\
                                                             A_{1h} & A_{1a} & A_{11} & A_{12} \\
                                                              A_{2h} & A_{2a} & A_{21} & A_{22} 
                                                          \end{array} \right) \,, \q {\bf{X}} = \left( \begin{array}{c}
                                                          H \\
                                                           A \\
                                                            D_1 \\
                                                            D_2 
                                                          \end{array} \right)   \,.  \label{linearA} \ee
The different terms $A_{ij}$ in $\tA$, which are functions of $\gamma$ and the non-dimensional parameters $R$, $S$, $k_h$, $b_h$, $k_a$ and $b_a$ governing the problem,  are given in Appendix \ref{app_coeffA}.

For a set of values of the  non-dimensional parameters, the system yields non-trivial solutions for eigenvalues $\gamma= k + i \sigma$ satisfying
\be \det [\tA(\gamma)] = 0 \,. \label{detA0} \ee
If the imaginary part $\sigma$ is negative, the system is unstable, and it will undergo oscillations with a non-dimensional frequency given by the real part $k$ of the eigenvalue $\gamma$.  On the contrary, if $\sigma >0$, any small perturbation in $z_s$ will be damped. Of special physical relevance is the characterization of the neutral hyper-surface in the parameter space where $\sigma=0$, which separates the stable from the unstable region, providing the flutter, or critical, velocity $U^*$ in terms of the different parameters, and also the corresponding  natural frequency $k=k_n$.

This flutter stability analysis extends that performed in \cite{ferna22}, which only considered the first flexural mode $d_1(t)$ (with $d=24 d_1$ in the notation of that reference). 
Consequently, the eigenvalues with  $D_2=0$, i.e., the solutions of   
\be \det [\tA_3(\gamma)]=0\,, \q\q \text{with} \q\q   \tA_3 =   \left( \begin{array}{ccc}
                                                             A_{hh} & A_{ha} & A_{ha}  \\
                                                            A_{ah} & A_{aa} & A_{a1}  \\
                                                             A_{1h} & A_{1a} & A_{11}     \end{array} \right) \,, \label{A3}
\ee
coincide with those described in that reference. We shall see that the instability results differ   substantially as the stiffness parameter $S$ decreases, because new unstable modes appear associated to the second deflection mode $d_2$, i.e., to the second natural frequency of the system, not considered in the previous work.                                                     

Before characterizing the parametric regions where the foil is unstable it is of interest to consider some particular cases and validate the results with some previous numerical ones.

\subsection{Some particular cases. Validation}
\label{sec_valida}
For a rigid foil, that is, for $S \to \infty$, one gets $d_1, d_2 \to 0$, and the eigenvalues $\gamma$ are obtained from 
\be \det[\tA_r] =   \det \left[ \begin{array}{cc}
                                                            A_{hh} & A_{ha}  \\
                                                            A_{ah} & A_{aa} 
                                                           \end{array} \right] = 0 \,. \label{knr}
\ee 
According to the solutions of this equation one obtains the  well known result  that  flutter instabilities  can only exist (solutions with $\sigma <0$ are only possible) for coupled pitch and heave passive motions of the rigid foil in some parametric ranges \citep[e.g.,][]{dowel15}. Neglecting the effect of the fluid-structure interaction and in absence of dampers ($b_a=b_h=0$), Eq.  \eqref{knr} yields the natural frequencies (with $\sigma=0$)
\be k_{nr0}= \sqrt{\frac{3k_a +2k_h \pm \sqrt{9 k_a^2+6k_a  k_h+4k_h^2}}{R}} \,. \label{knr0}
\ee      
More particularly,  when only pitching motion is allowed, i.e. for $k_h \to \infty$  so that $H \to 0$, the equation $A_{aa}=0$ yields, for a rigid foil in absence of dampers, the well known result (corresponding to $k_{nr0}^-$ in the limit $k_a/k_h \to 0$)
\be k_{nr0a}=\sqrt{\frac{3 k_a}{2 R}} \,.
\ee
Similarly, for $k_a \to \infty$, so that $A \to 0$ and only passive heave is allowed, the resulting equation $A_{hh}=0$ yields the well known result (limit $k_h/k_a \to 0$ of $k_{nr0}^-$)
\be k_{nr0h}=\sqrt{\frac{k_h}{R}} \,.
\ee

Of more interest for the validation of the present model is the opposite configuration of a flexible foil, but clamped at its leading edge; i.e., with both $k_a,k_h \to \infty$, so that   passive pitch and heave     are inhibited and flutter instabilities are only associated to the flexural deformation of the foil. The eigenvalues are given by      
\be \det[\tA_f] =   \det \left[ \begin{array}{cc}
                                                            A_{11} & A_{12}  \\
                                                            A_{21} & A_{22} 
                                                           \end{array} \right] = 0 \,. \label{knf}
\ee       
In vacuum (i.e., neglecting the contributions of the fluid force and moments), this equation yield the first two natural frequencies  of a clamped beam       
\be k_{nf0}= \sqrt{\frac{7}{953} \left( 629 \pm 2 \sqrt{88189} \right) \frac{S}{R} } \,, \label{knf0}
\ee     
which reproduce very accurately their exact values, especially the first one, with relative errors of $0.006\%$ and $5.7\%$, respectively \citep{ferna25}. 

\begin{figure}
\centerline{\epsfig{file=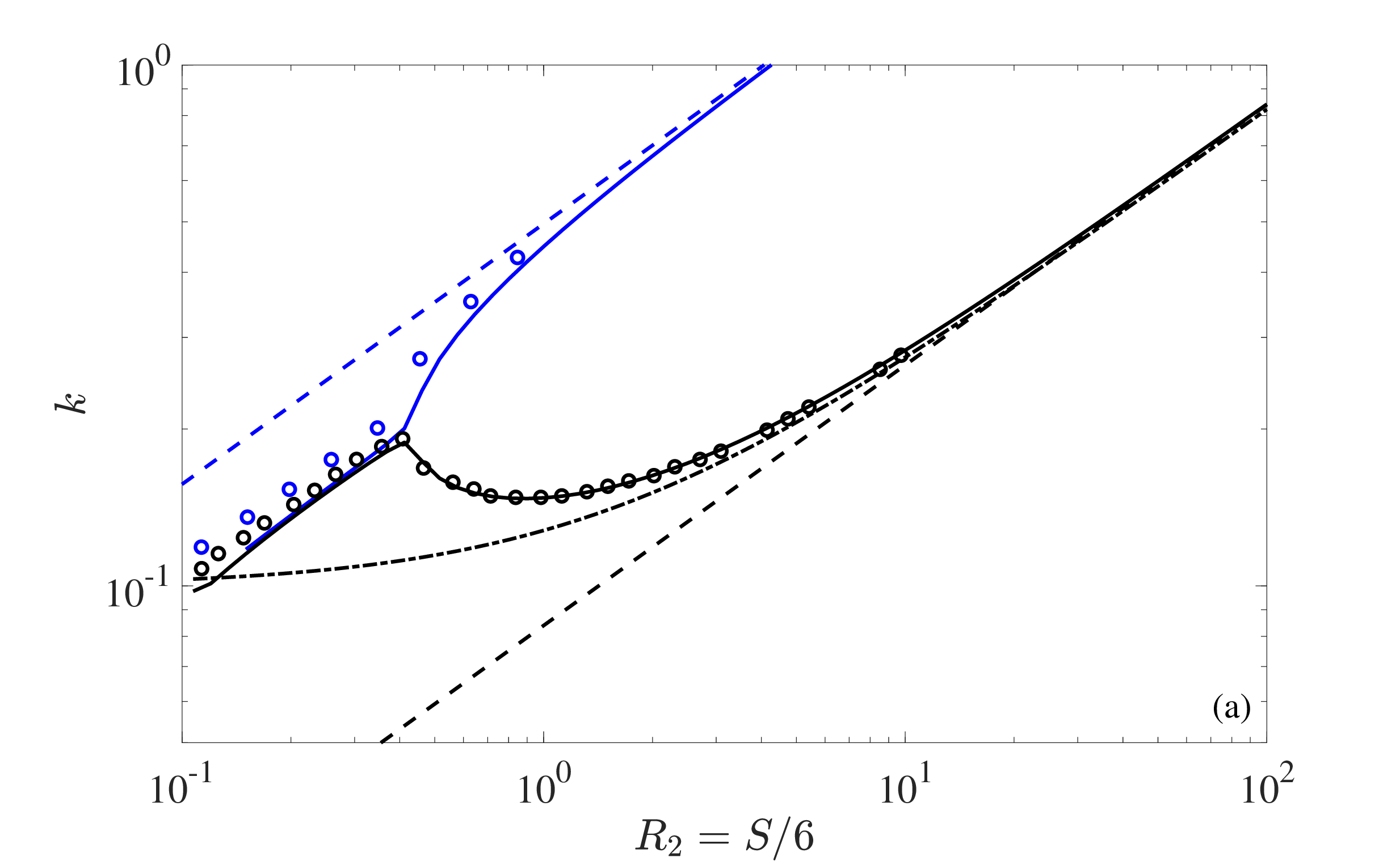,width=.55\linewidth}\epsfig{file=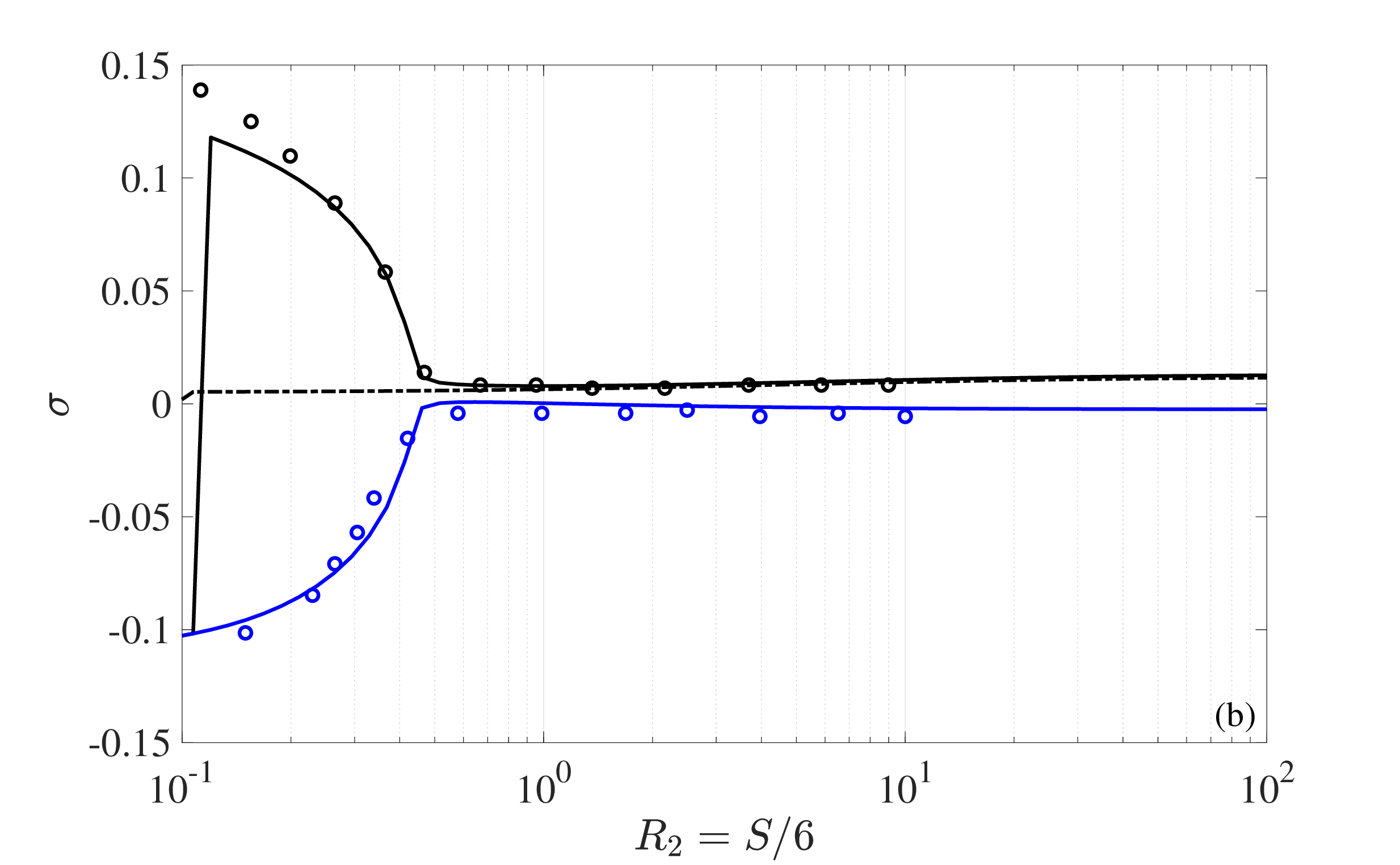,width=.55\linewidth}}
  \caption{Comparison of the real (a) and the  imaginary part (b) of the eigenvalue $\gamma$ resulting  from \eqref{detA0} (continuous lines) for a clamped foil ($k_a,k_h \to \infty$) with the exact values obtained by Alben \cite{alben08a} (circles) for $R=219.4$ ($R_1=109.7$ in Alben's notation) as $S$ ( $= 6 R_2$) is varied. Also plotted in (a) with dashed lines are the natural frequencies in vacuum \eqref{knf0}, and with dash-and-dotted lines the real and imaginary parts of the eigenvalues from \eqref{A3}, when only the first flexural mode is considered ($d_2=0$).}
\label{compara_Alben}
\end{figure}

\begin{figure}
\centerline{\epsfig{file=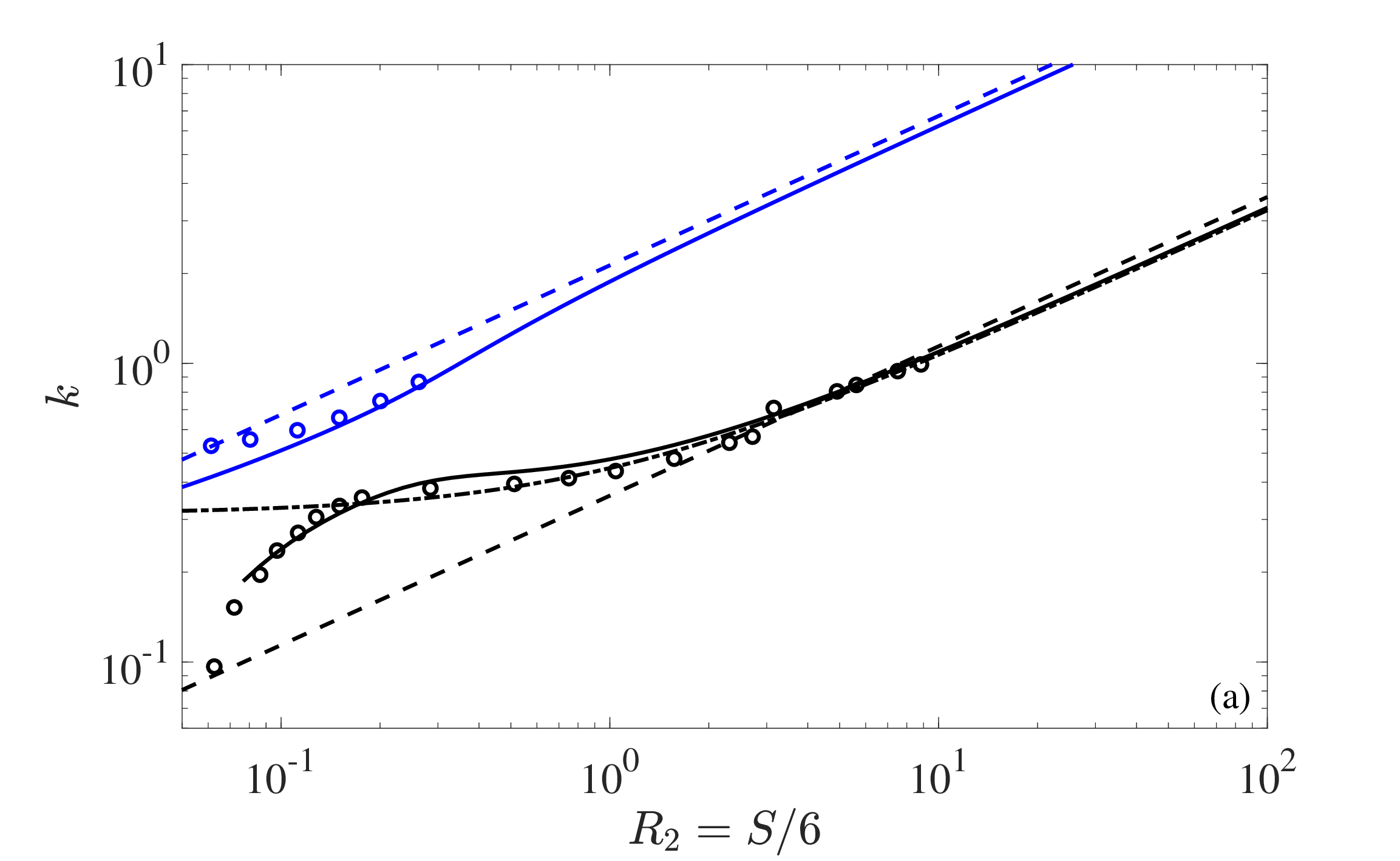,width=.55\linewidth}\epsfig{file=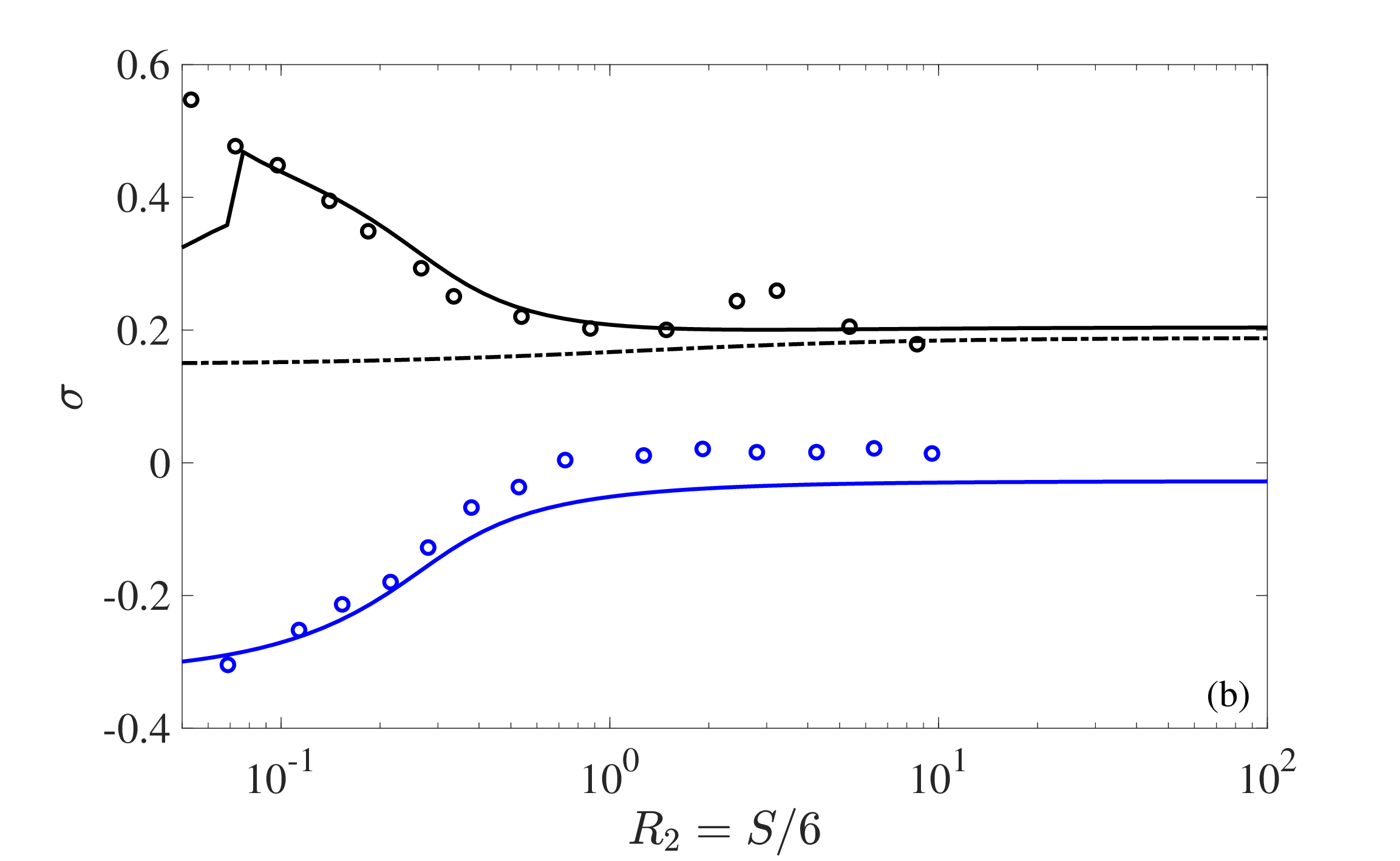,width=.55\linewidth}}
  \caption{As in Fig. \ref{compara_Alben}, but for  for $R=11.898$ ($R_1=5.949$).}
\label{compara_Albena}
\end{figure}

Similar accuracy is obtained when one takes into account the effect of the fluid. To check that, Figs. \ref{compara_Alben} and \ref{compara_Albena} compare the eigenvalues obtained from \eqref{knf} with the exact ones obtained numerically by Alben \cite{alben08a} for the first couple of eigenvalues  for two particular values of $R$, as $S$ is varied. The cases selected are those  reported in Figs. 4e-4h in \cite{alben08a} for $R=219.4$ and $R=11.898$ ($R_1=R/2=109.7$ and $5.949$ in Alben's notation), with the real and imaginary parts of $\gamma = k + i \sigma$  plotted in terms of $R_2$, which corresponds to $S/6$ in the present notation. For each value of $R$ and $S$, the nonlinear equation \eqref{detA0} with $k_a,k_h \to \infty$ (or, equivalently, Eq. \eqref{knf})  is solved  for $\gamma$ starting from the corresponding eigenvalues \eqref{knf0} in vacuum, which  are also plotted in Figs. \ref{compara_Alben} and  \ref{compara_Albena} with dashed lines for reference sake (they both approach asymptotically for $S \to \infty$).

It is observed that the present approximation  reproduces very accurately the natural modes with lower frequency (plotted with black lines in the figures), correcting the previous results that only contemplated the first mode of deformation, obtained here from Eq. \eqref{A3} and also plotted with dash-and-dot lines. The differences between them grow substantially as the stiffness decreases,  more pronouncedly for the growth rate than for the frequency. In fact,  $\sigma$ is drastically unpredicted by the previous approach for  $R_2 \lesssim 0.5$ (or $S \lesssim 3$), specially after the bifurcation of $\sigma$ as $S$ decreases. On the other hand, the agreement of the present second natural mode (with higher frequency, plotted with blue lines  in the figures)  with the exact one is also quite good. Notice that the present results are valid till so low a stiffness as $S \approx  0.3$, well below the value of $S$ where the frequencies of the two natural mode approach each other and their grow rates bifurcate into a clear stable mode ($\sigma >0$) and a clear unstable one ($\sigma <0$). The previous approach with just a single flexural mode (dashed-and-dotted lines) became  poor for $S \lesssim  10$, and capturing only the lower frequency, stable mode. Thus, the present approach with two flexural modes not only extends the region of validity from $S$ of order $10$ to $S$ of order $10^{-1}$, but also captures the unstable mode which was missing in the previous analytical approach for this clamped foil. This will be very relevant for the characterization of the parametric ranges for flutter  instabilities. Of course, all the other bunch of modes appearing for very flexible foils with $S \lesssim 10^{-1}$ (see Fig. 4 in \cite{alben08a}) are not covered by the present analytical approach, failing to predict the flutter characteristics for very small values of $S$, specially when $R$ is also small, as we shall see below.

Further validations of the present analytical results against other available numerical results will be reported in sections \ref{sec_clamp_comp} and \ref{sec_pinned_comp} below.

\begin{figure}
\centerline{\epsfig{file=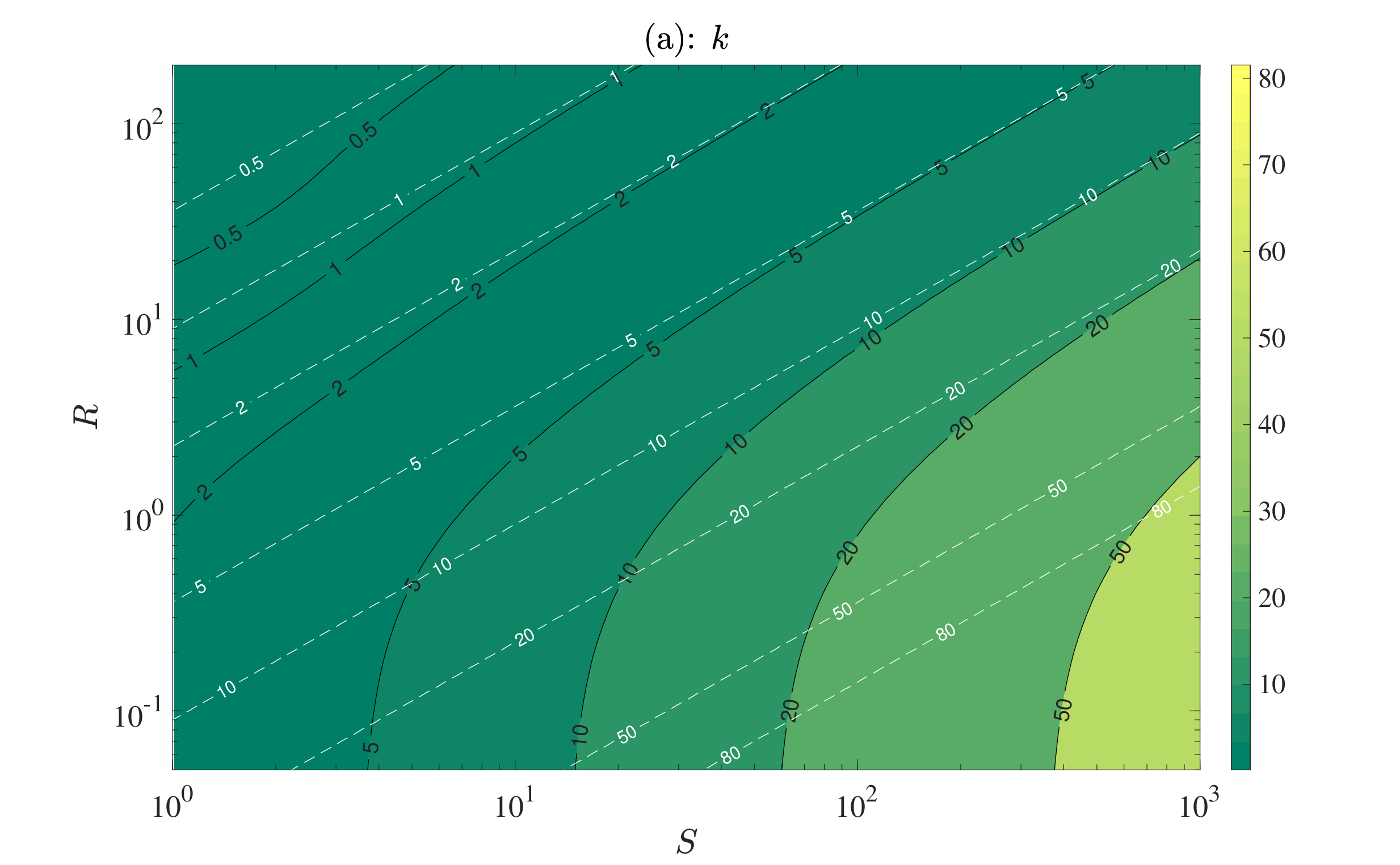,width=.55\linewidth}\epsfig{file=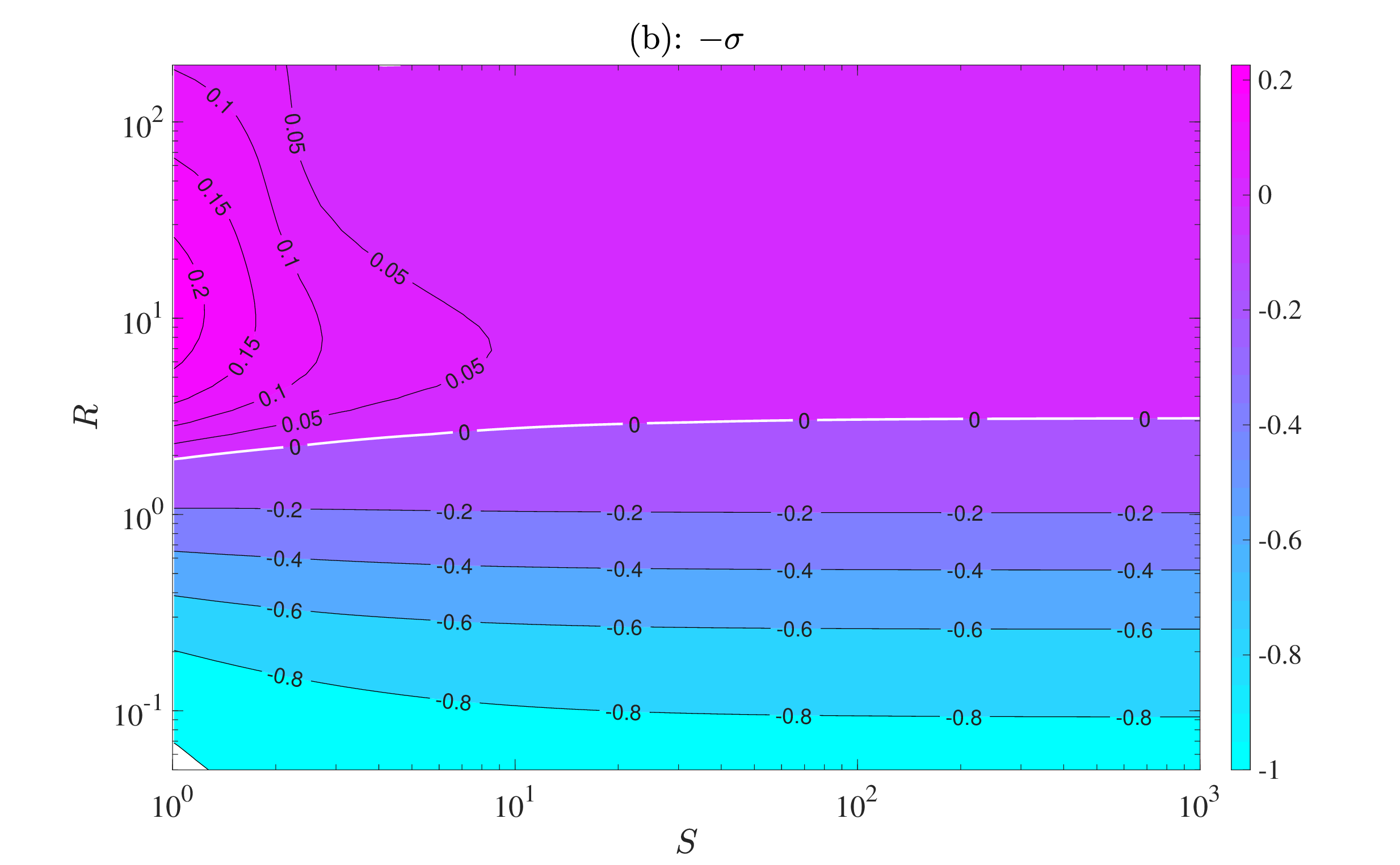,width=.55\linewidth}}
  \caption{Real (a) and minus imaginary (b) part  of $\gamma=k+i \sigma$ resulting from \eqref{detA0}, when $k_{nf0}^+$ from \eqref{knf0} is used as seed, as $S$ and $R$ are varied for $k_a, k_h \to \infty$. Some values of $k_{nf0}^+$ are plotted in (a) with dashed lines, and the neutral curve $\sigma=0$ is plotted  in (b) with a thick white line (note that $-\sigma$ is the growth rate).}
\label{Fig_res1}
\end{figure}

\section{Results and discussion}
\label{sec_results}
To map the eigenvalues $\gamma$ as the different non-dimensional parameters are varied, the complex algebraic equation \eqref{detA0}  is solved  using Matlab's function \textit{fsolve}  starting from the corresponding natural frequency when the FSI is neglected, which are  given by \eqref{knr0} of by \eqref{knf0}. We start with the clamped foil case, extending the results in Figs. \ref{compara_Alben} and \ref{compara_Albena} to all values of the mass ratio $R$. Unless otherwise stated, all the reported results are computed with  damping constants $b_h=b_a=0.5$ (see \S \ref{sec_dampers} for the effect of the dampers).

\begin{figure}
\centerline{\epsfig{file=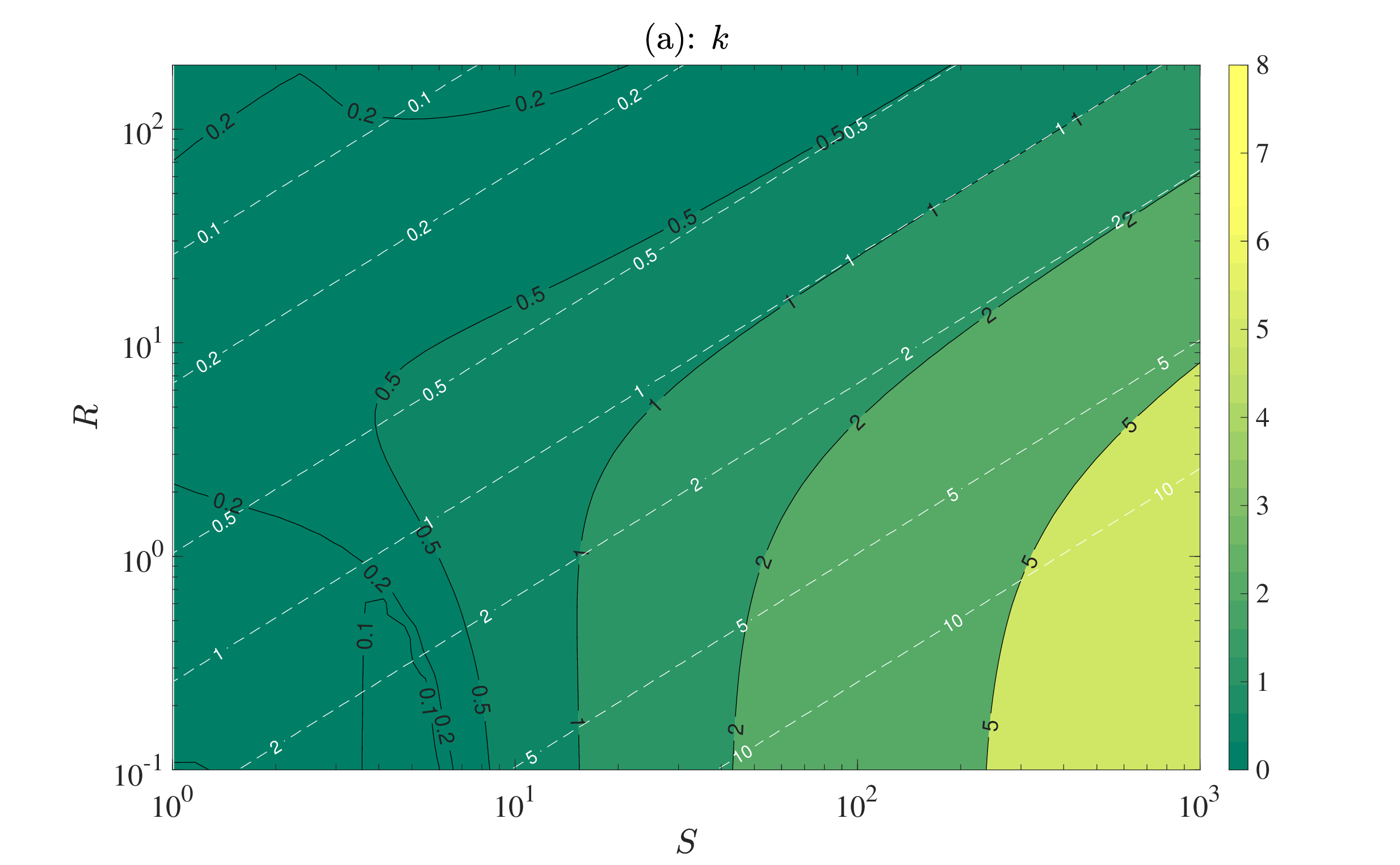,width=.55\linewidth}\epsfig{file=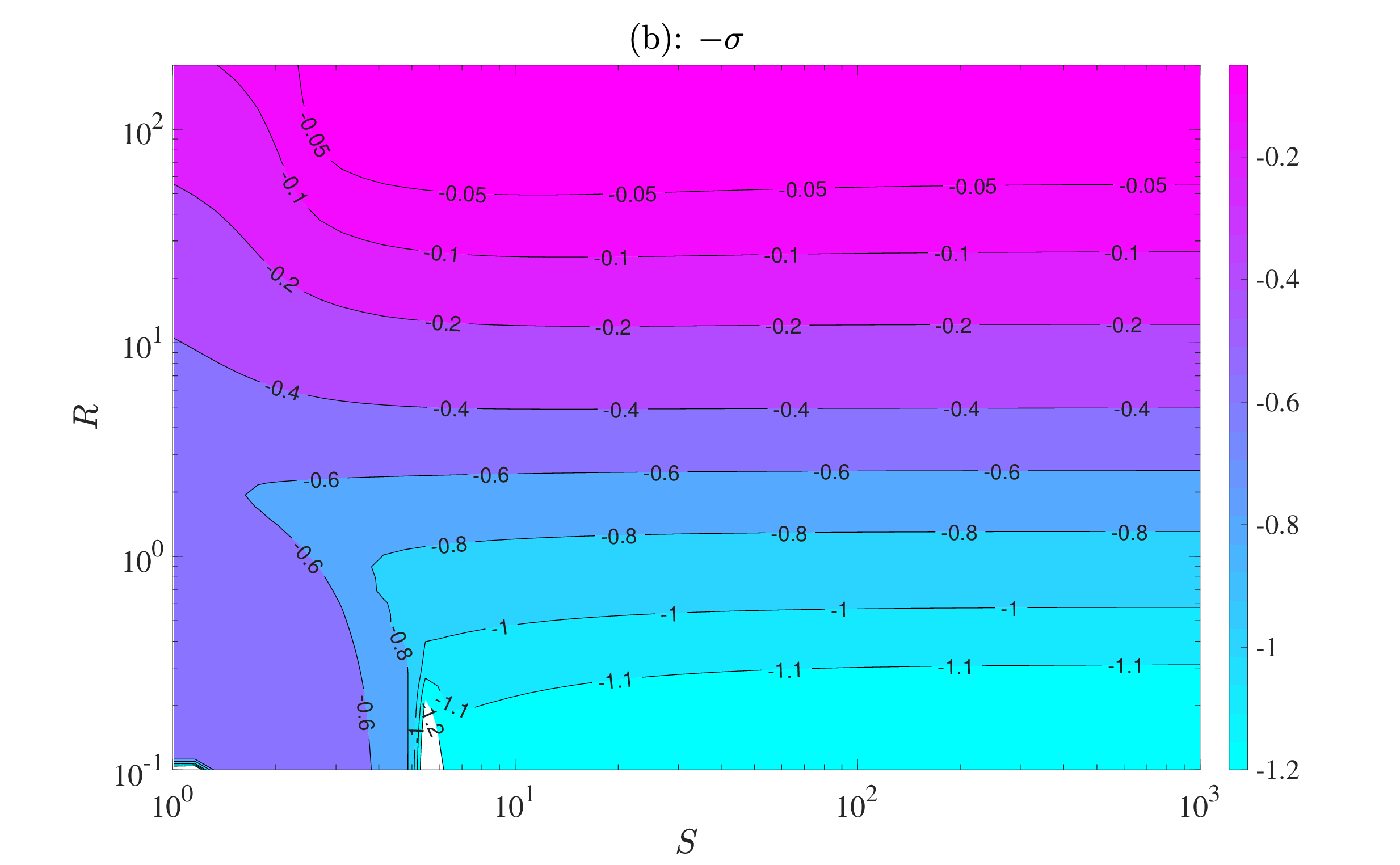,width=.55\linewidth}}
  \caption{As in Fig. \ref{Fig_res2} but for $k_{nf0}^-$.}
\label{Fig_res2}
\end{figure}

\subsection{Clamped flexible plate}

Figure \ref{Fig_res1} shows the real and imaginary parts of $\gamma$ for a clamped foil (i.e., when $k_a, k_h \to \infty$) on the $(S,R)-$plane corresponding to the higher frequency mode, which is obtained by solving Eq. \eqref{detA0} starting from  $k_{nf0}^+$ given by Eq. \eqref{knf0} with positive sign. In agreement with Figs. \ref{compara_Alben} and \ref{compara_Albena}, Fig. \ref{Fig_res1}(b) shows that the foil is unstable for all values of $S$ when $R=219.4$ and $R=11.898$ (notice that now the growth rate $-\sigma$ is plotted, so that flutter instabilities correspond to positive values in the figure). As $S$ is varied, Fig. \ref{Fig_res1}(b) tells us that flutter instabilities are generated only above a critical value of the mass ratio $R$ that depends on the stiffness $S$, marked with a thick white line. However, the growth rate is very small for large $S$, increasing suddenly when $S$ decreases below around ten.   
A maximum growth rate is reached about $R=11$ for small $S$. Figure \ref{Fig_res1}(a) shows the corresponding natural reduced frequencies $k$ (we omit the subindex $n$ in $k_n$ for simplicity in the notation), together with some  values of $k_{nf0}^+$ (i.e., without taking into account the effect of the fluid) with dashed lines, which asymptotically coincide with $k$ for large $R$ and $S$. The frequency corresponding to the most unstable case shown with $R \approx 11$ for $S$ around unity is $k \approx 0.65$.

The eigenvalues corresponding to the lower frequency mode, obtained by solving Eq.  \eqref{detA0} with  $k_{nf0}^-$ from Eq. \eqref{knf0} as a seed, are plotted in Fig. \ref{Fig_res2}. No flutter instabilities are observed, with the foil-fluid system becoming more estable as $R$ decreases, like in the higher frequency mode.

\begin{figure}
\centerline{\epsfig{file=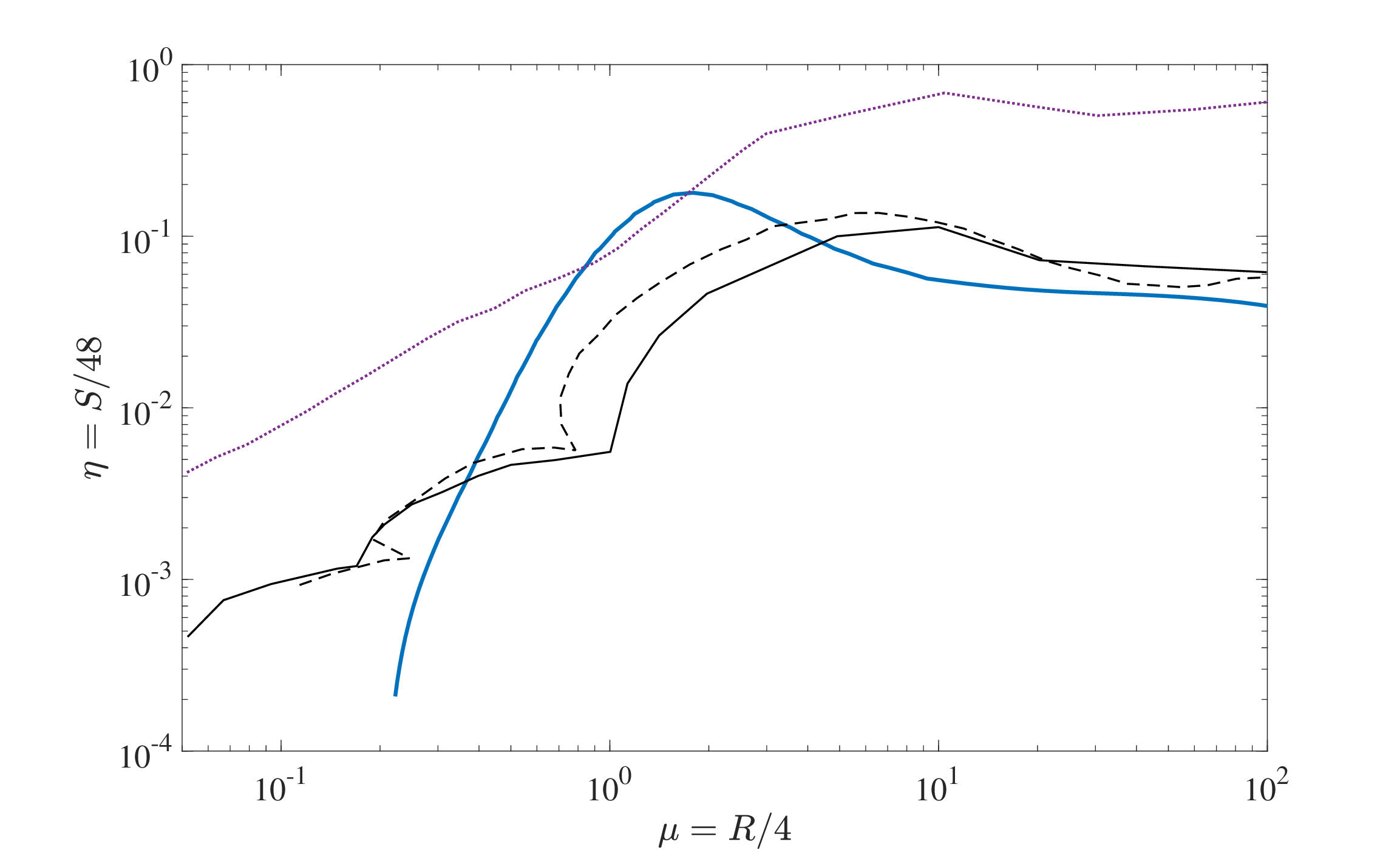,width=.75\linewidth}}
  \caption{Comparison of the curve for $-\sigma=0.05$ in the $(R,S)-$plane for a clamped plate (thick blue) with  the critical curve for the stability obtained by Michelin et al. \cite{micll08} (black solid), by Eloy et al. \cite{elola08} (dashed) and by Alben and Shelley \cite{albsh08} (dotted), as reported in  Fig. 4(b) of Ref. \cite{micll08}.}
\label{compara_Michelin}
\end{figure}

\subsubsection{Comparison with previous results} \label{sec_clamp_comp}

Figure \ref{compara_Michelin} compares  the present analytical results for the critical curve for the stability of a flexible plate clamped at the leading edge with some previos numerical results reported in Fig. 4 of Michelin eta al. \cite{micll08}. The results are plotted   in the $(R,S)-$plane but in the notation of Ref. \cite{micll08}, where $\mu=R/4$ is the mass ratio and $\eta=S/48$ is the stiffness. In particular, the present results  corresponding to $-\sigma=0.05$ are compared with the critical curve for the stability of the plate obtained by Michelin et al. \cite{micll08} using a train of unsteady point of vortices to model the vortical wake. Also included in the figure are previous numerical results from a linear stability analysis  by Eloy et al. \cite{elola08}, and from  a vortex sheet approach by Alben and Shelley \cite{albsh08}, all of them reported in Fig. 4(b) of Ref. \cite{micll08}. In fact, Eloy et al. \cite{micll08}  extended and improved the former  linear stability analysis of Kornecki et al. \cite{kordo76} and validated the numerical  results  with experimental data. 

As expected from the comparison  made in \S \ref{sec_valida} for the eigenvalues in this clamped-free  case, the present analytical results agree reasonably well with the numerical ones when $S \gtrsim 0.1$, corresponding in the critical curve to $R \gtrsim 1$. For smaller values of $S$ and $R$, a train of unstable flapping flag  modes appears that are not captured by the present analysis.

\subsection{Passive heave with no pitch}

We now explore the flutter stability of the flexible foil as the linear spring constant $k_h$ decreases from infinity to allow for heaving motion, but with the pitch still inhibited ($k_a \to \infty$), as the stiffness $S$ is varied for relevant values of the inertia parameter $R$. In particular we select $R=10$ as a representative case of a foil in air, which is unstable for a clamped plate, and $R=0.2$, representative of a foil in water, stable when clamped according to the present model.

\begin{figure}
\centerline{\epsfig{file=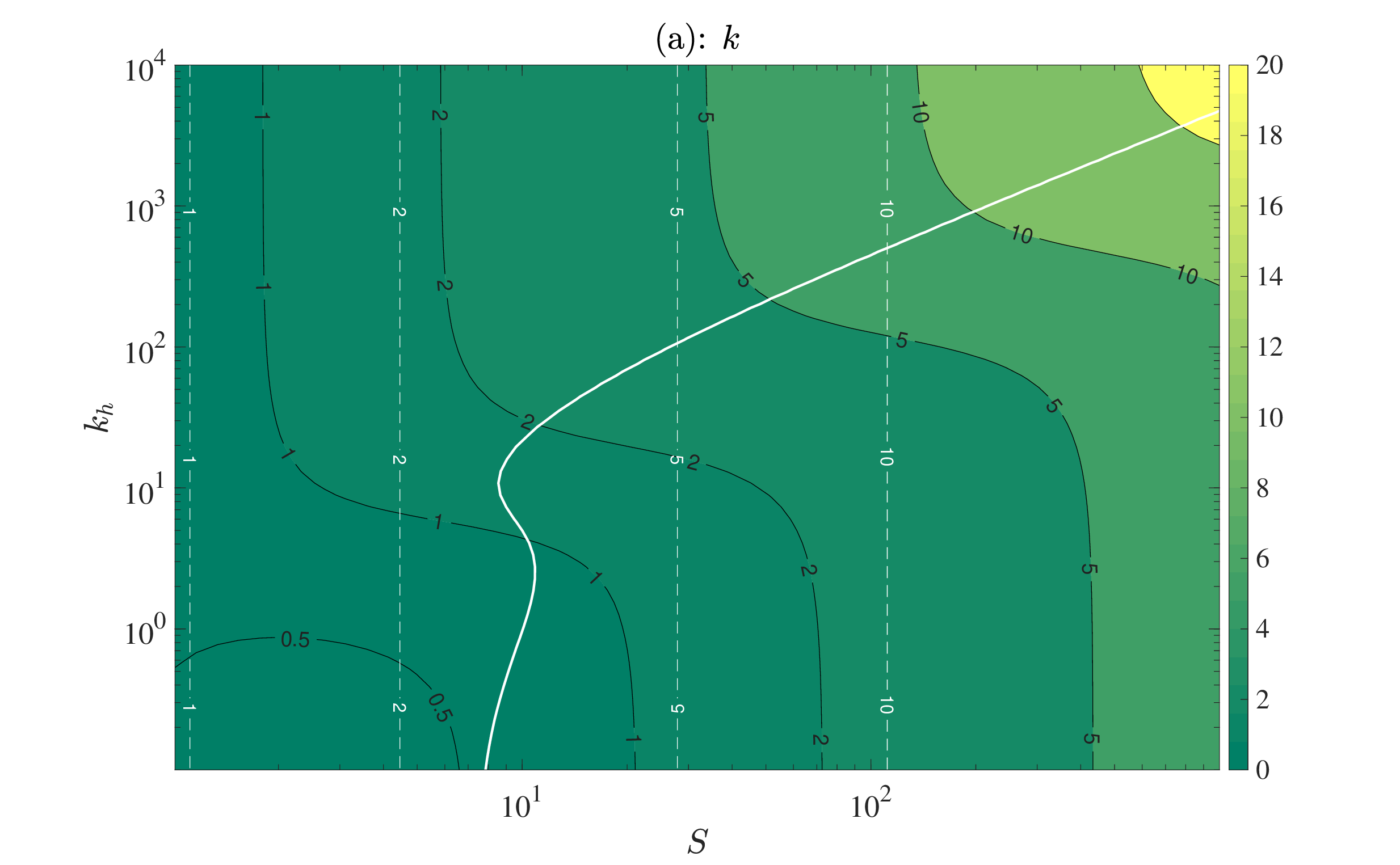,width=.55\linewidth}\epsfig{file=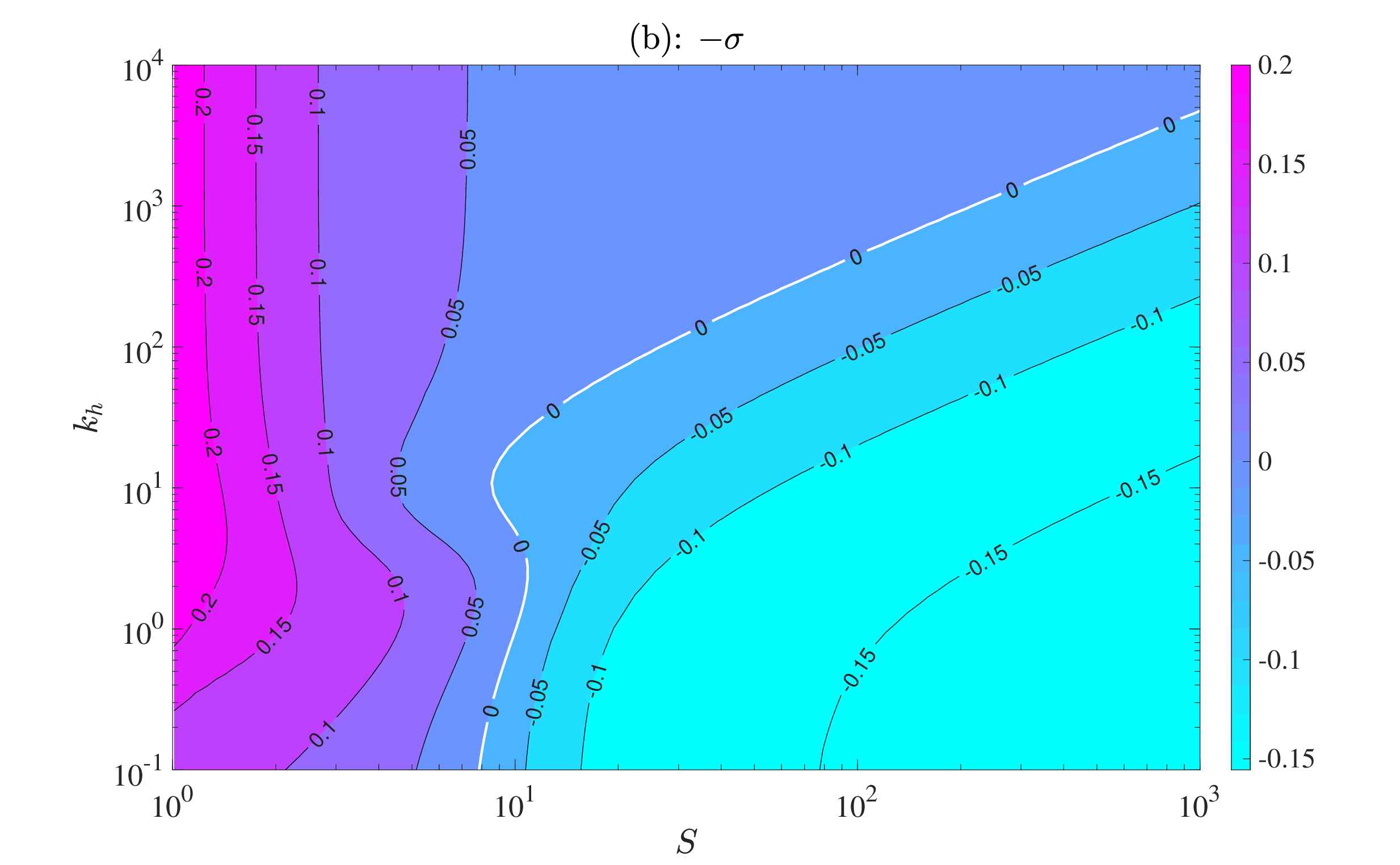,width=.55\linewidth}}
  \caption{Real (a) and minus imaginary (b) part  of $\gamma=k+i \sigma$ resulting from \eqref{detA0} for $R=10$ when $k_{nf0}^+$ from \eqref{knf0} is used as seed, as $S$ and $k_h$ are varied for $k_a \to \infty$. Some values of $k_{nf0}^+$ are plotted in (a) with dashed lines, and the neutral curve $\sigma=0$ is plotted with a thick white line.}
\label{Fig_res3}
\end{figure}

Figure \ref{Fig_res3} shows the eigenvalues in the $(S,k_h)-$plane for $R=10$ resulting from the unstable flexural mode (i.e., resulting from $k_{nf0}^+$ and tending to those of  a clamped foil plotted in Fig. \ref{Fig_res1} as $k_h \to \infty$). It is observed that, for a finite value of $k_h$,  the unstable flexural mode becomes stable for stiffness $S$ above a threshold value which depends on $k_h$, $S^*(k_h)$ say, marked by the neutral stability curve (white line) in Fig. \ref{Fig_res3}. $S^*(k_h)$ decreases from infinity from a clamped plate (when $k_h \to \infty$) until $k_h$ is of the order of ten, and then remains almost constant with $S$ also about ten. Actually, for $k_h$ above of about $10^2$, the threshold stiffness decays almost linearly,  and the unstable mode for $S<S^*$ corresponds to a pure flexural mode, with the natural frequencies $k$ practically independent on $k_h$ (almost vertical contour lines in Fig. \ref{Fig_res3}(a), closely following the purely flexural frequencies $k_{nf0}^+$, also plotted with dashed lines in that figure). Below $k_h \approx 10^2$, the effect of the spring mode becomes important (the natural frequencies depend on both $k_h$ and $S$) and the growth rate increases slightly. For small $k_h$, the instabilities are  again dominated by the flexural mode, with natural frequencies almost independent of $k_h$, but  now they are related to  the second flexural mode associated to $k_{nf0}^-$ plotted in Fig. \ref{Fig_res2},  which is stable for large $k_h$, but becomes unstable for small $k_h$ (actually, this lower-frequency flexural mode is the same found unstable in \cite{ferna22}, which is the only one present when $d_2=0$, as discussed above). For intermediate values of $k_h$ the spring mode dominates, with almost horizontal contour lines in Fig. \ref{Fig_res3}(a), so that the natural frequencies depend more on $k_h$ than on $S$. 

No flutter  instabilities are found when $R=0.2$ for any value of $k_h$ or $S$, as it will also become clear with the analysis described just below when $R$ is varied for a given $k_h$.

\begin{figure}
\centerline{\epsfig{file=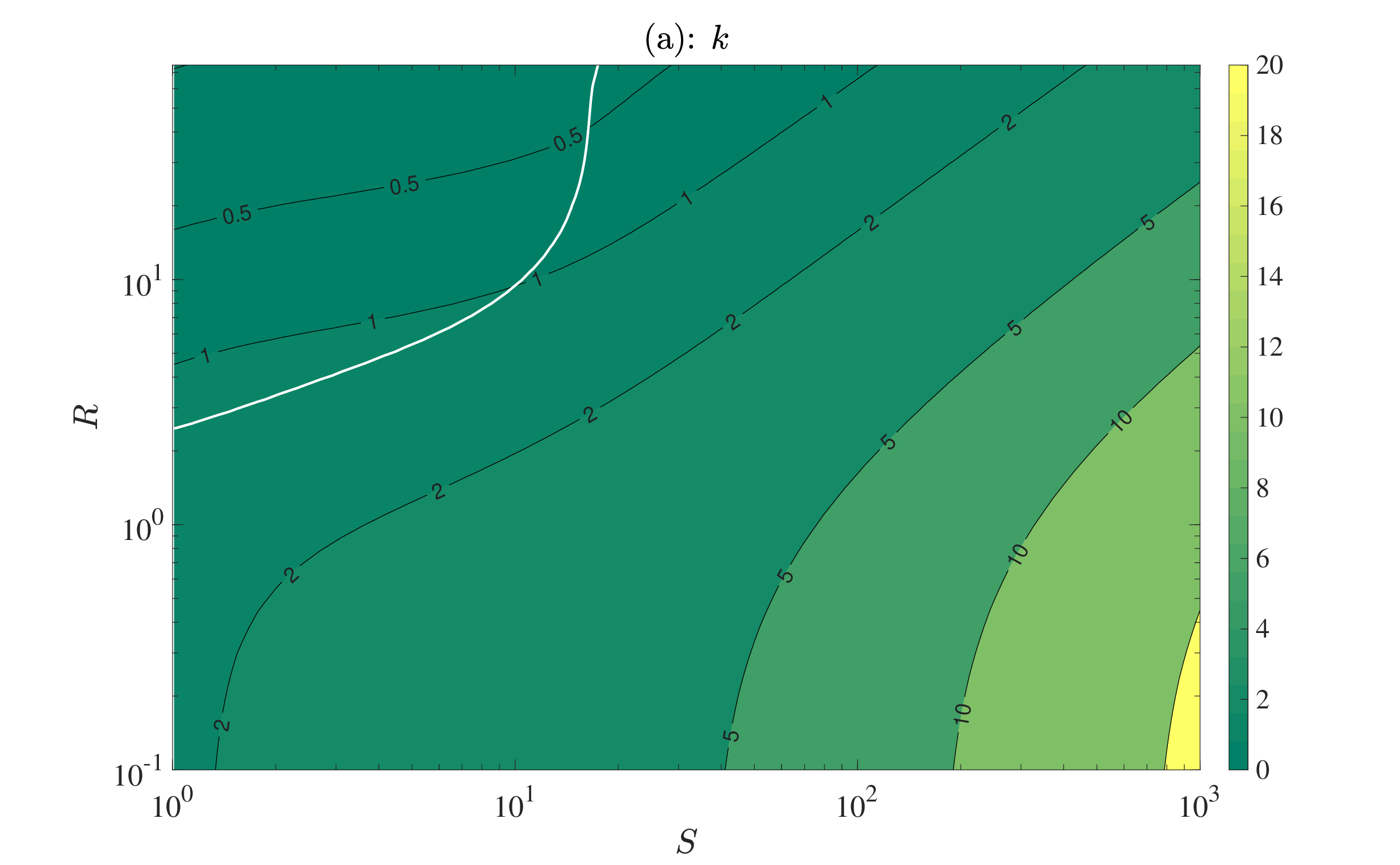,width=.55\linewidth}\epsfig{file=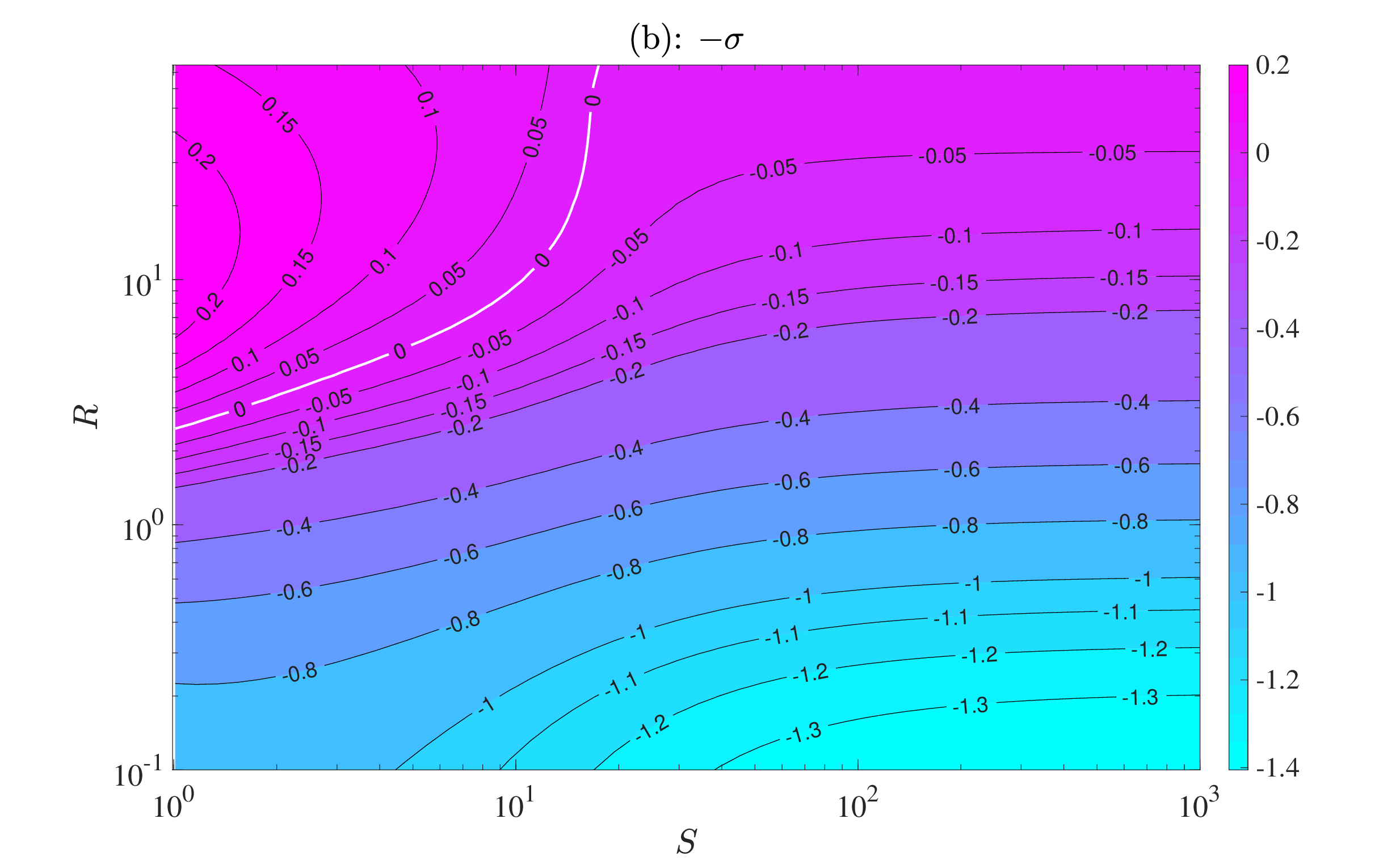,width=.55\linewidth}}
  \caption{Real (a) and minus imaginary (b) part  of $\gamma=k+i \sigma$ resulting from \eqref{detA0} for $k_h=4$ when $k_{nf0}^+$ from \eqref{knf0} is used as seed, as $S$ and $R$ are varied for $k_a \to \infty$.  The neutral curve $\sigma=0$ is plotted with a thick white line.}
\label{Fig_res4}
\end{figure}

The modes plotted in Fig. \ref{Fig_res3} are the only unstable ones for $R=10$ without passive pitch. It is clear in the figure that no flutter instabilities are generated in a rigid foil ($S \to \infty$) when one allows passive heave ($k_h$ below a threshold value) without pitch, which is the classical result that two degrees of freedom, heave and pitch, are required for flutter instability of a rigid plate \citep{theod35,dowel15}. Figure \ref{Fig_res3} shows with a white line the threshold value of $k_h^*(S)$ below which there is no instability for heave only for $R=10$ (alternatively, the threshold stiffness $S^*(k_h)$ above which there is no flutter). Obviously,  this threshold curve depend on $R$. In addition, it is also known that these flutter instabilities with just passive heave of a flexible plate are only possible for large enough values of $R$ \citep{ferna22}. Actually, as already shown in Figs. \ref{Fig_res1} and \ref{Fig_res2}, pure flexural flutter instabilities are only possible above a threshold curve $R^*(S)$ when $k_h \to \infty$. This threshold function varies with $k_h$. Figure \ref{Fig_res4} shows it for $k_h=4$, a value of $k_h$ that according to Fig. \ref{Fig_res3}(b) roughly corresponds  to the maximum growth rate when $R=10$. The instabilities above the neutral curve $R^*(S)$ in Fig. \ref{Fig_res4} are mixed spring-flexural modes for this relatively low value of $k_h$.  In particular,  $R^*$ varies between $R\simeq 2.45$ for $S=1$ and asymptotes to $R \to \infty$ for $S \approx 20$, so that no flutter instabilities are possible for $S \gtrsim 20$ when $k_h=4$ with no passive pitch. As $k_h$ increases, this upper stiffness limit for flutter instabilities grows monotonically as shown above, while the lower limit of $R$ for each $S$ tends to the asymptote given by the white curve in Fig. \ref{Fig_res1}(b).

\begin{figure}
\centerline{\epsfig{file=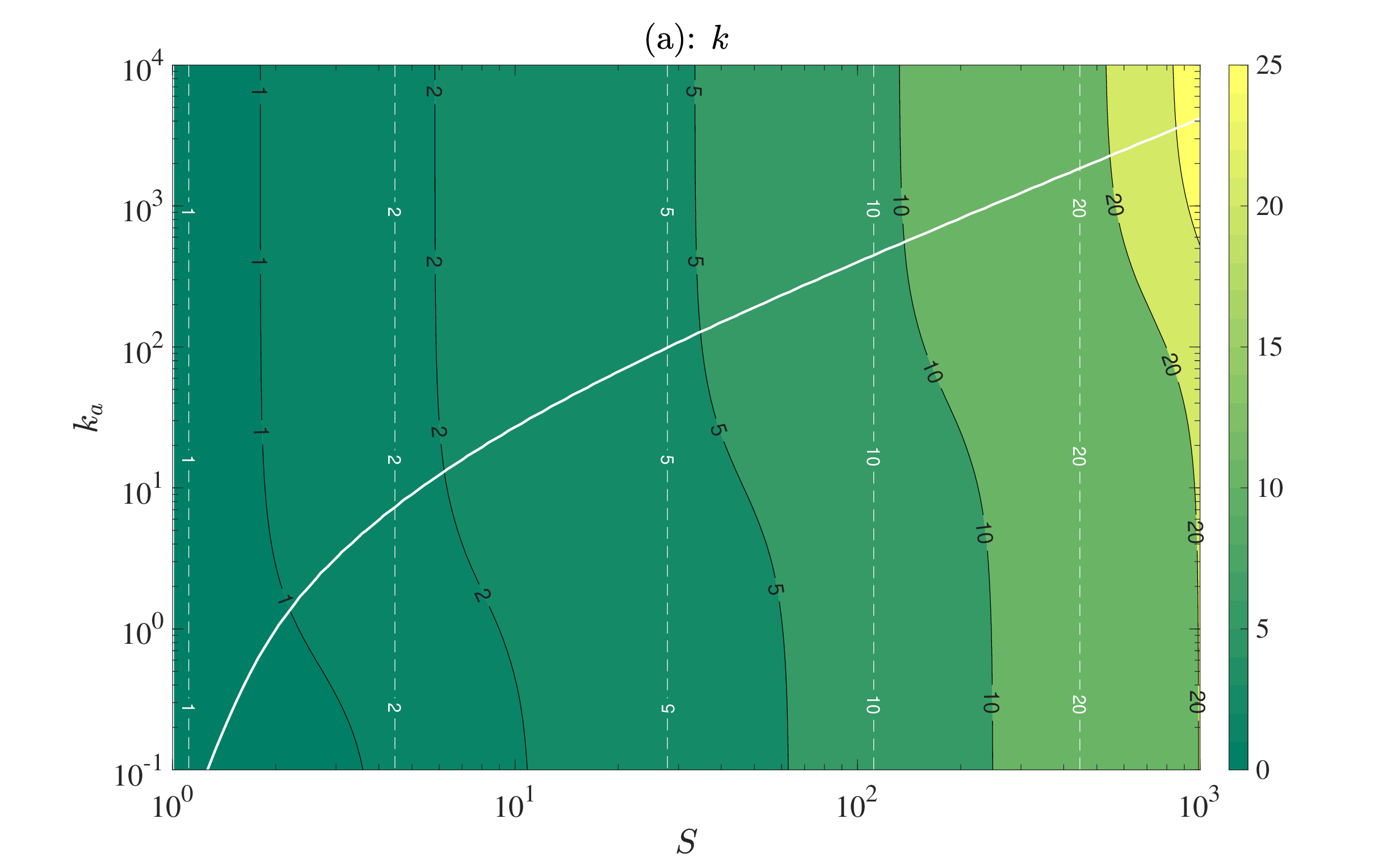,width=.55\linewidth}\epsfig{file=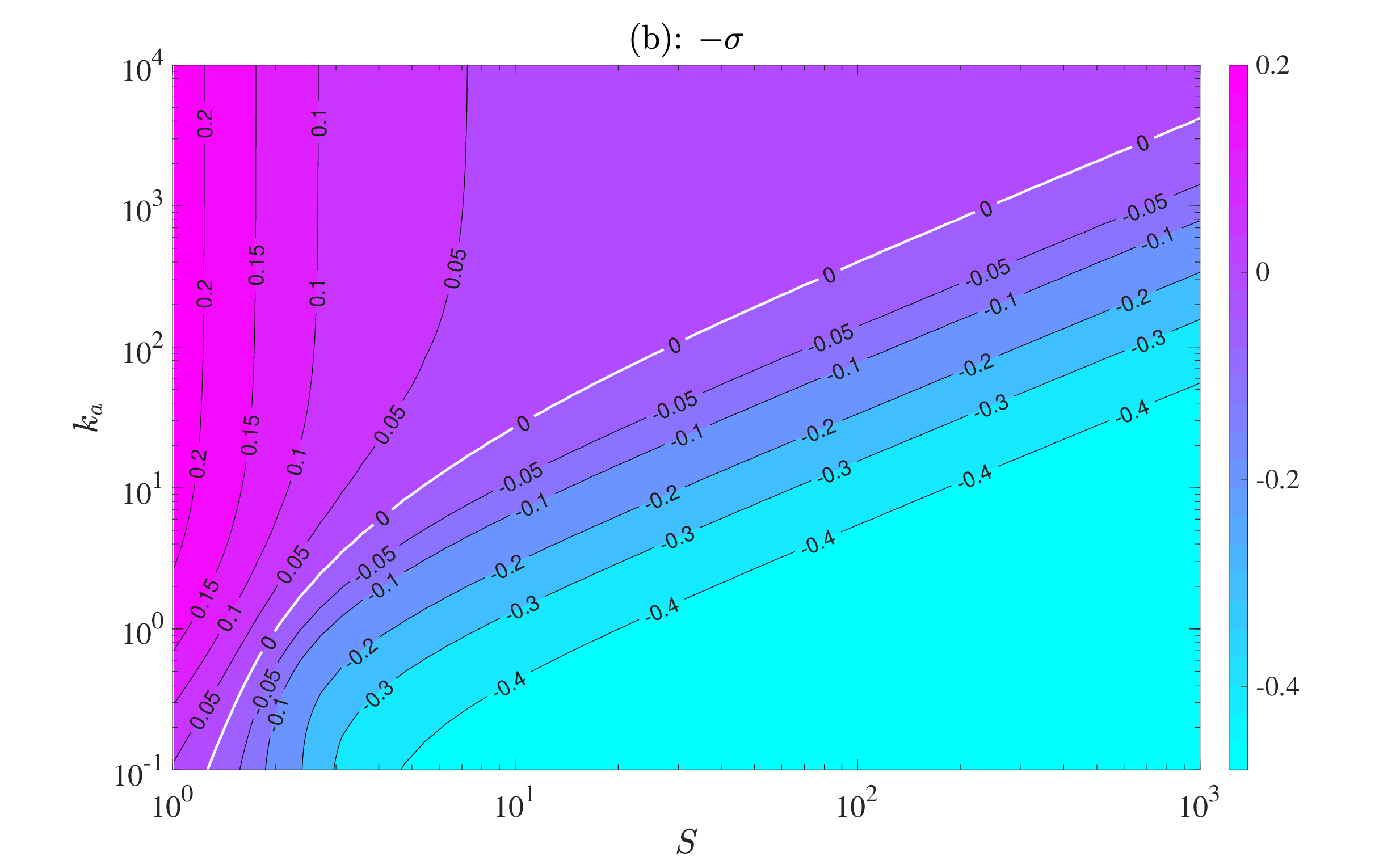,width=.55\linewidth}}
  \caption{Real (a) and minus imaginary (b) part  of $\gamma=k+i \sigma$ resulting from \eqref{detA0} for $R=10$ when $k_{nf0}^+$ from \eqref{knf0} is used as seed, as $S$ and $k_a$ are varied for $k_h \to \infty$. Some values of $k_{nf0}^+$ are plotted in (a) with dashed lines, and the critical curve for instability ($\sigma=0$) is plotted with a thick white line.}
\label{Fig_res5}
\end{figure}

\subsection{Passive pitch with no heave}

We now analyze the effect of passive pitch (varying the torsional spring constant $k_a$)  in the flutter stability of a flexible plate when the heave is inhibited ($k_h \to \infty$).

Figure \ref{Fig_res5} shows the results in the $(S,k_a)-$plane for $R=10$. As in the case with only passive heave plotted in Fig. \ref{Fig_res3}, the stiffness range of the clamped  foil flexural instability decreases rapidly as $k_a$ decreases from infinity. But, as a difference, in addition that this range of instability decreases much more quickly, no enhancement of the flexural  instability due to the torsional spring does appear for values of $k_a$  of order unity: the flutter instability is basically a flexural instability mode throughout all the values of $k_a$, as evidenced by the natural frequencies shown in Figure \ref{Fig_res5}(a). As $k_a$ decreases, no substantial pitching motion is generated and the original flexural instability simply decays very quickly with  $k_a$. Since this flexural instability is associated to the deformation mode $d_2(t)$, as shown in the previous sections, this result is in agreement with those reported in \cite{ferna22}, where only the deformation mode $d_1(t)$ was considered and no flutter instabilities with pitch-only motion were found for a flexible plate.

\begin{figure}
\centerline{\epsfig{file=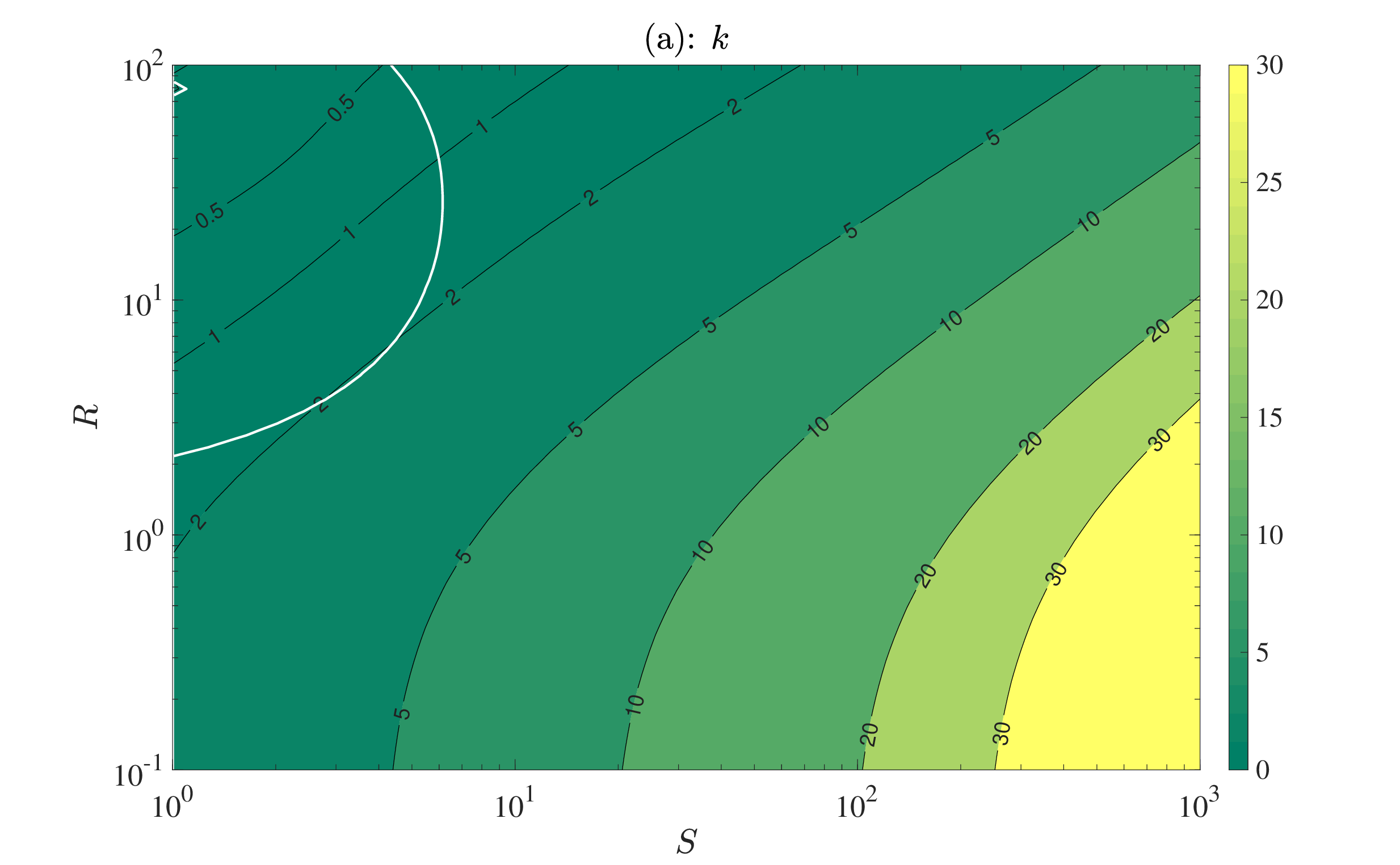,width=.55\linewidth}\epsfig{file=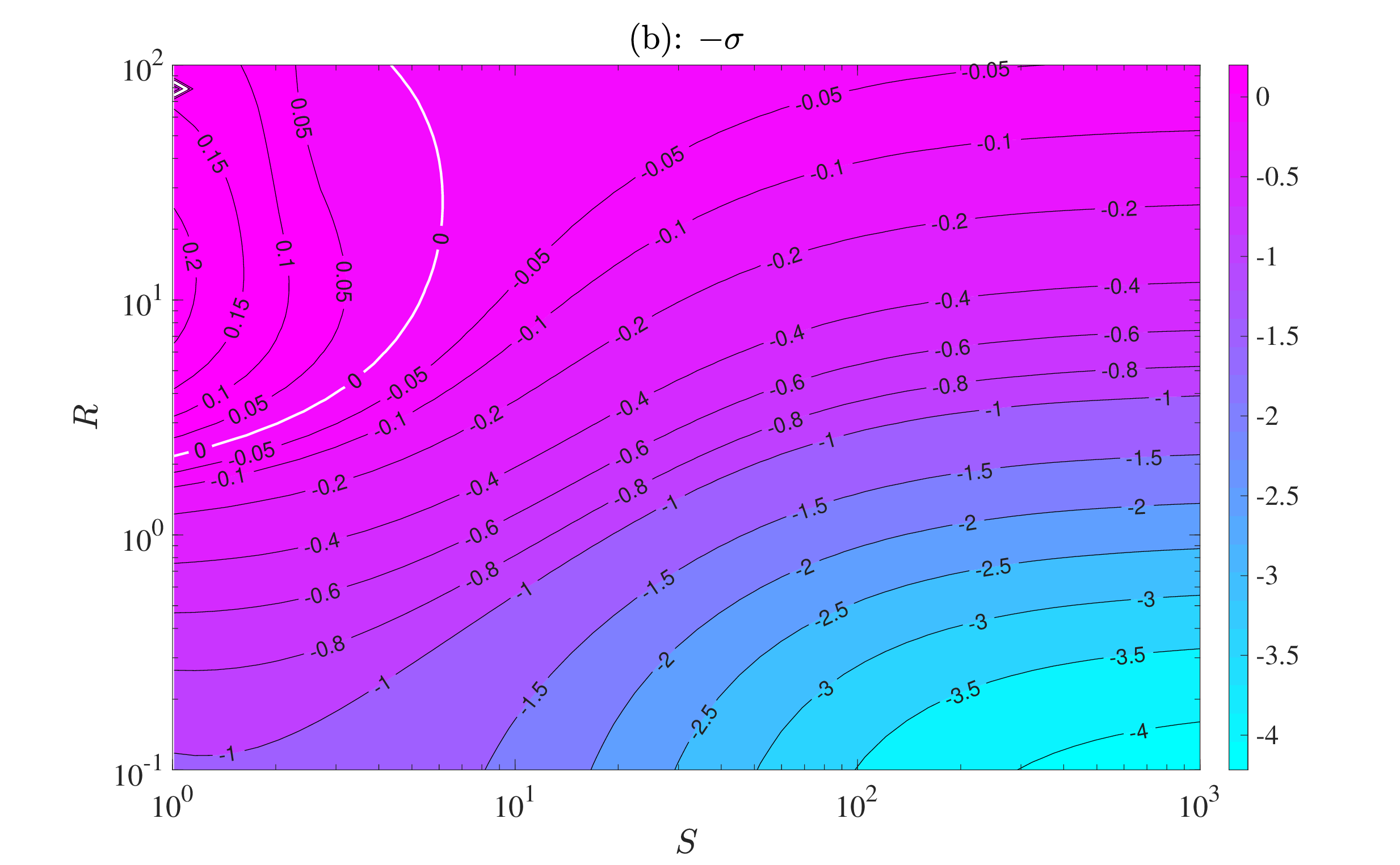,width=.55\linewidth}}
  \caption{Real (a) and minus imaginary (b) part  of $\gamma=k+i \sigma$ resulting from \eqref{detA0} for $k_a=10$ when $k_{nf0}^+$ from \eqref{knf0} is used as seed, as $S$ and $R$ are varied for $k_h \to \infty$.  The neutral curve $\sigma=0$ is plotted with a thick white line.}
\label{Fig_res6}
\end{figure}

As in the previous cases for a clamped foil and with only passive heave, the present flutter instabilities with only passive pitch are  generated above a threshold mass ratio $R^*$ which depends on $S$ and $k_a$. The case with $k_a \to \infty$ for a clamped plate is already plotted in Fig. \ref{Fig_res1}. To illustrate how this threshold $R^*(S)$ becomes more constrained as $k_a$ decreases, Fig. \ref{Fig_res6} shows it for $k_a=10$ with a white line.

\subsubsection{Comparison with previous results} \label{sec_pinned_comp}

\begin{figure}
\centerline{\epsfig{file=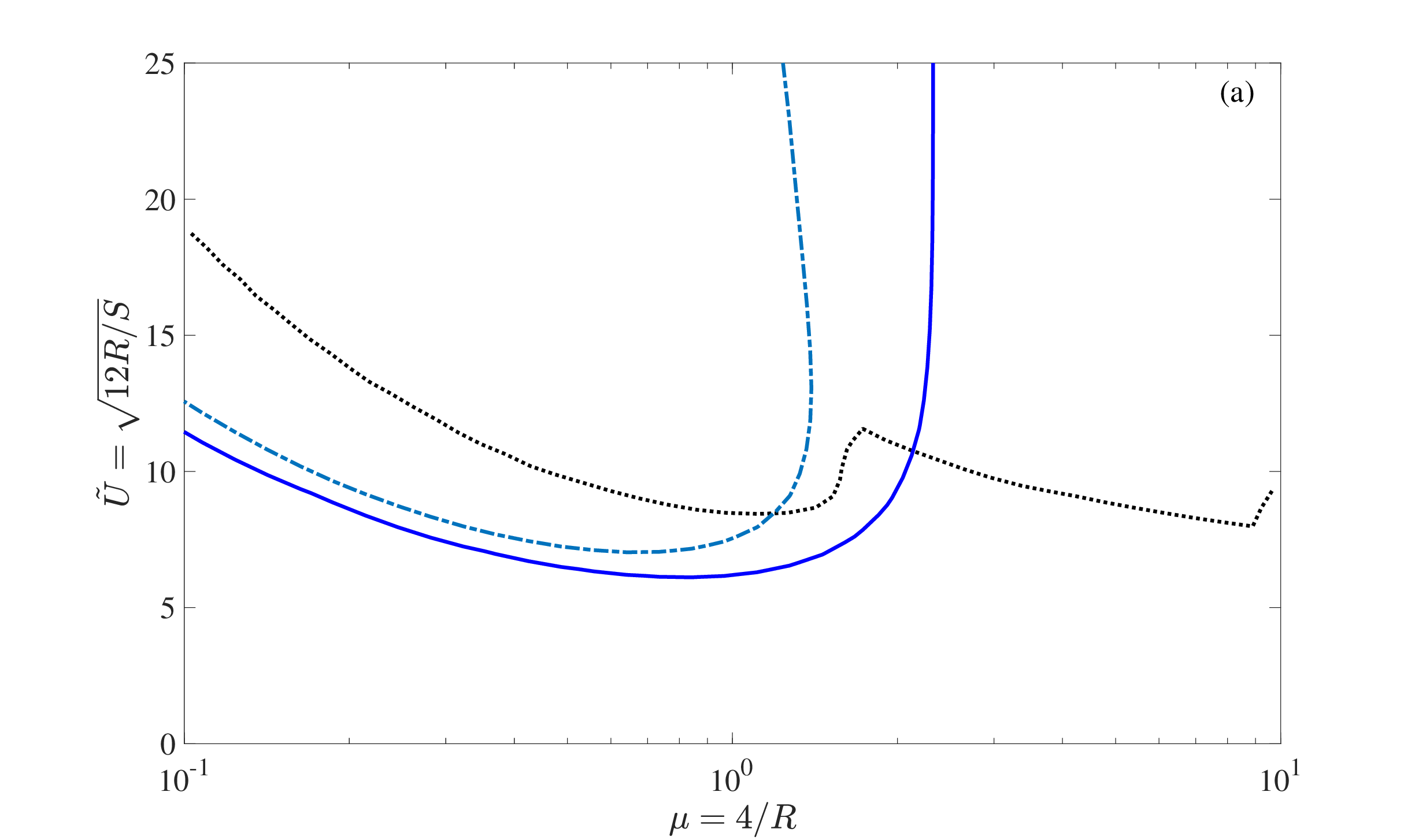,width=.55\linewidth}\epsfig{file=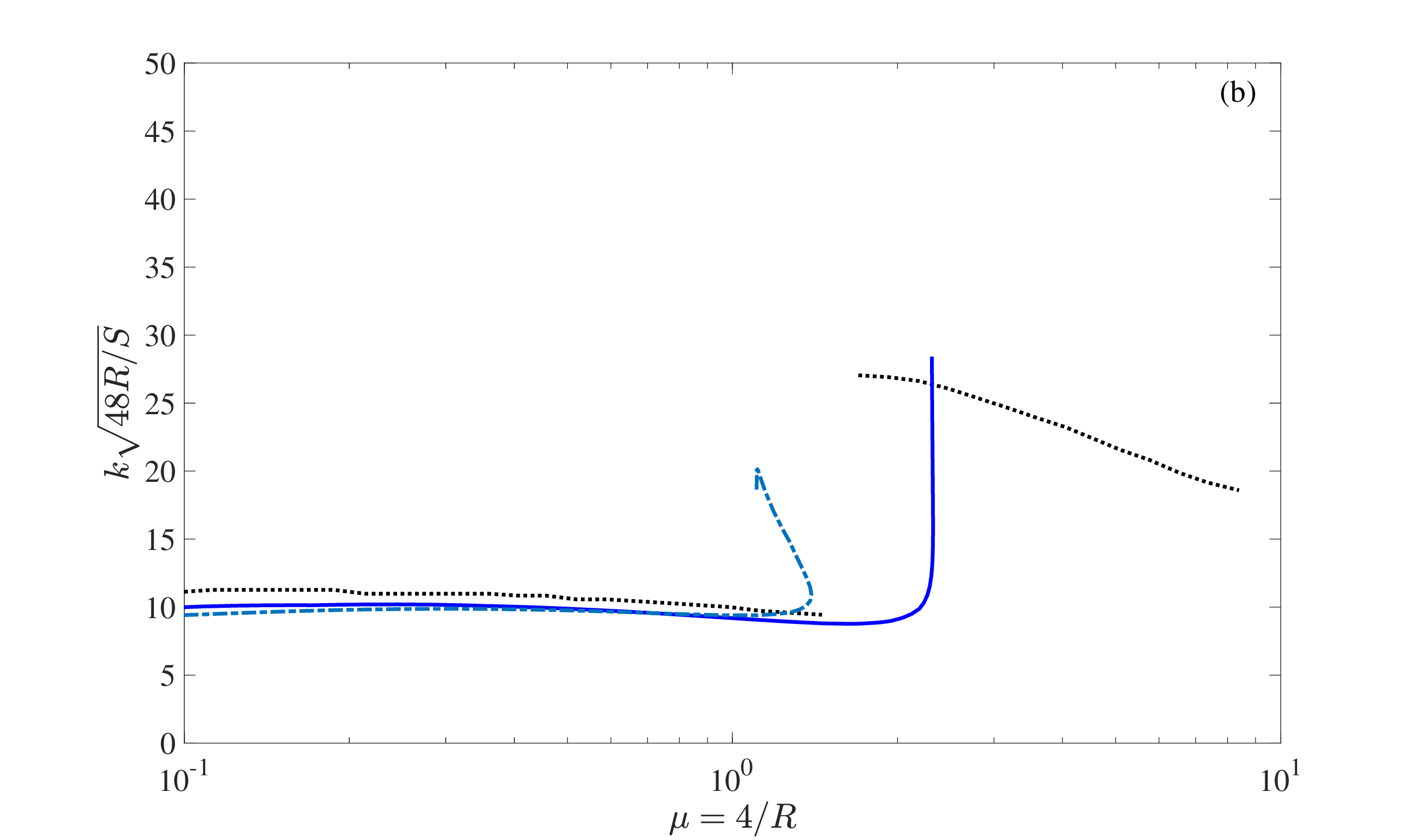,width=.55\linewidth}}
  \caption{Comparison of the non-dimensional critical  flutter velocity (a) and frequency (b) vs. the mass ratio obtained numerically by Chad Gibbs et al. \cite{chawa12} for a pinned-free beam with aspect ratio $0.5$ (dotted lines; form Fig. 8 in that reference) with the present results for $k_a \simeq 0$, $k_h \to \infty$, $b_a=b_h=0$, with $\sigma=0$ (solid lines) and $\sigma=-0.05$ (dash-and-dotted lines).}
\label{compara_Gibbs}
\end{figure}

The instability region in the $(R,S)-$plane for $k_a=10$ \textit{inside} the white neutral curve in the left upper corner of Fig. \ref{Fig_res6} decreases with $k_a$, but remains finite for $k_a=0$. The flutter stability of this  pinned-free flexible plate  has also been analyzed by a number of authors, corresponding to $k_a=0$ and $k_h \to \infty$ in the present configuration of an elastically mounted flexible plate.  

Figure \ref{compara_Gibbs} compares numerical results by Chad Gibbs et al. \cite{chawa12} for the non-dimensional critical  flutter velocity of a pinned-free beam with an aspect ratio (span divided by chord-length) of $0.5$ as a function of the mass ratio with the present analytical results for $k_a=0$ and $k_h \to \infty$. The mass ratio and the non-dimensional flutter velocity  defined by Chad Gibbs et al. are, in the present notation, $\mu = 4/R$ and $\tilde{U}=\sqrt{12 R/S}$, respectively.

Figure \ref{compara_Gibbs}(a) compares the non-dimensional flutter velocity reported in Fig. 8 of Ref. \cite{chawa12} with the present results for $\sqrt{12 R/S}$  for $\sigma=0$ and for $\sigma=-0.05$. Figure \ref{compara_Gibbs}(b) shows the corresponding non-dimensional frequency  to compare with the non-dimensional frequency reported in Fig. 8(b) of Chad Gibbs et al. \cite{chawa12}, which in the present notation is $k \sqrt{48 R / S}$. In spite of the finite  aspect ratio of the plate in the numerical results by Chad Gibbs et al. \cite{chawa12}, there is a reasonable good agreement of the lowest-frequency flutter mode with the present analytical results for a 2-D plate (infinite aspect ratio), specially for the frequency. But the main consequence that can be drawn from this comparison is that it provides a good picture of the validity range of the present analytical formulation already stated and checked in several places above: it only recovers the lowest frequency unstable modes and their corresponding growth rate, and therefore their corresponding critical flutter velocity. These flutter instabilities are the only ones  when the stiffness parameter $S \gtrsim 10^{-1}$,  which can develop flutter instabilities for mass ratios $R$ larger than a critical value of order unity.

\subsection{Coupled pitch-heave-deformation flutter instabilities}

Before analyzing the effect of the flexibility of the foil, we briefly explore the well-known coupled pitch-heave  flutter  of a rigid foil using the present formulation.

\begin{figure}
\centerline{\epsfig{file=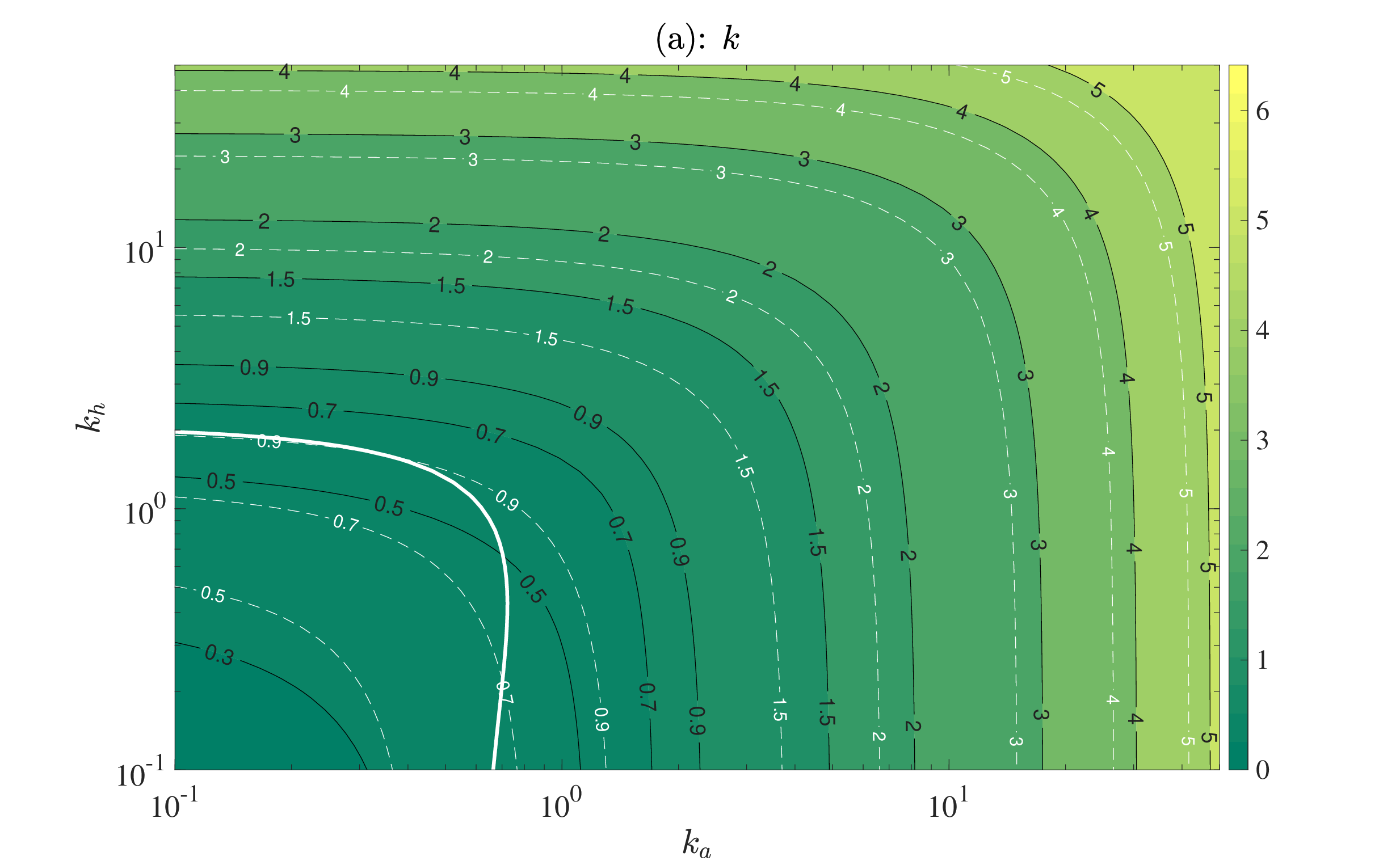,width=.55\linewidth}\epsfig{file=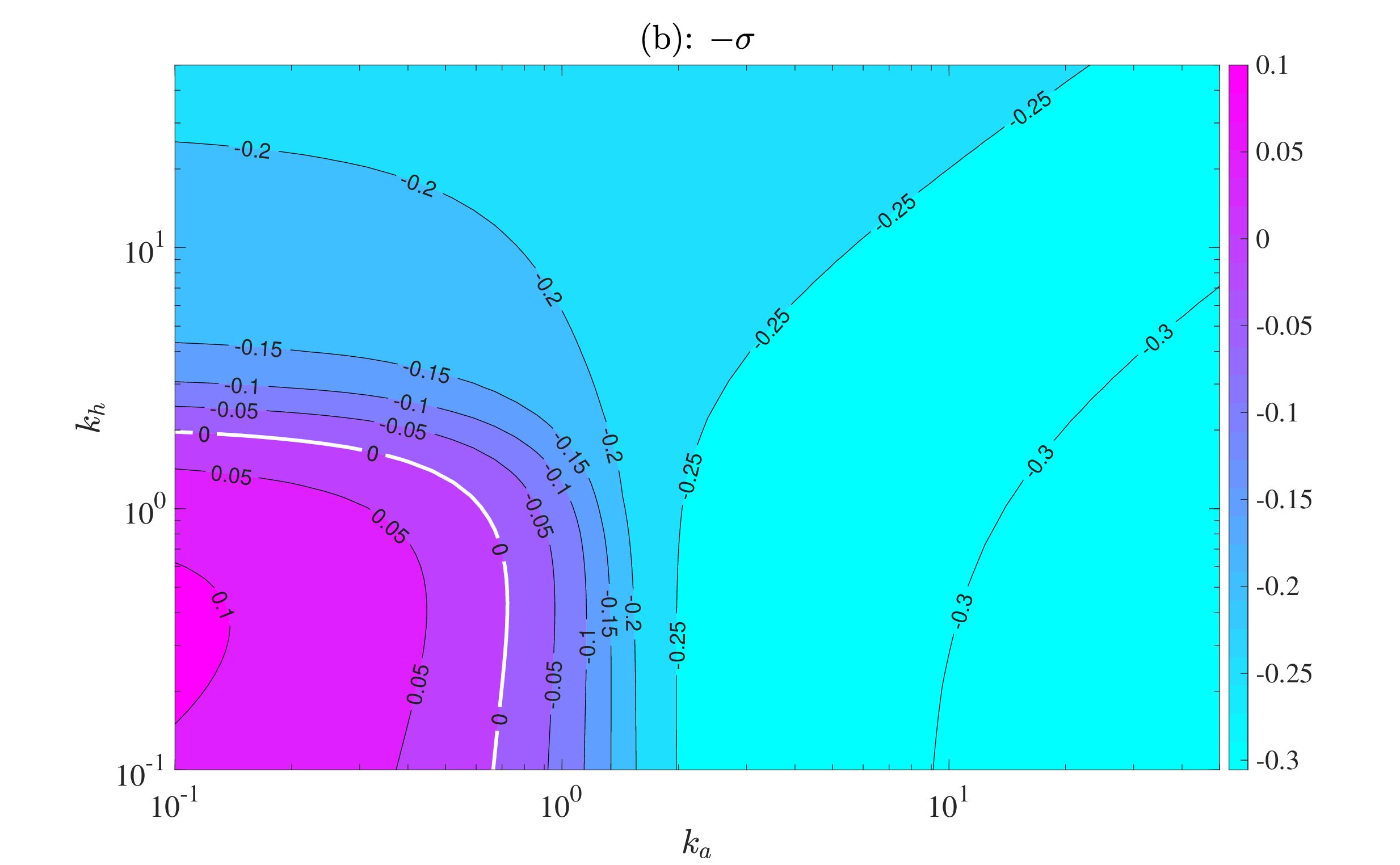,width=.55\linewidth}}
  \caption{Real (a) and minus imaginary (b) part  of $\gamma=k+i \sigma$ for a \textit{rigid} foil with $R=10$, resulting from \eqref{knr} when $k_{nr0}^+$ from \eqref{knr0} is used as seed, as $k_h$ and $k_a$ are varied.  Some values of $k_{nr0}^+$ are plotted in (a) with dashed lines, and the neutral curve $\sigma=0$ is plotted with a thick white line.}
\label{Fig_res7}
\end{figure}

Figure \ref{Fig_res7} maps on the $(k_a,k_h)$-plane the eigenvalues resulting from Eq. \eqref{knr} [i.e., for a rigid foil, or equivalently setting $S \to \infty$ in \eqref{detA0}] for a mass ratio $R=10$ when the natural frequency in vacuum $k_{nr0}^+$ (positive sign in \eqref{knr0}) is used as a seed. Some values of $k_{nr0}^+$ are also plotted in Fig. \ref{Fig_res7}(a) with dashed lines, approaching the natural frequencies that take into account the FSI as $k_a$ and $k_h$ increase. Figure \ref{Fig_res7}(b) shows that the flutter instabilities for this spring mode (associated to the coupling between the linear and torsional springs) are located in  a region of the plane with both $k_h$ and $k_a$ of order unity and below. No flutter instabilities associated to the other natural frequency in vacuum $k_{nr0}^-$ (negative sign in \eqref{knr0}) are found.

\begin{figure}
\centerline{\epsfig{file=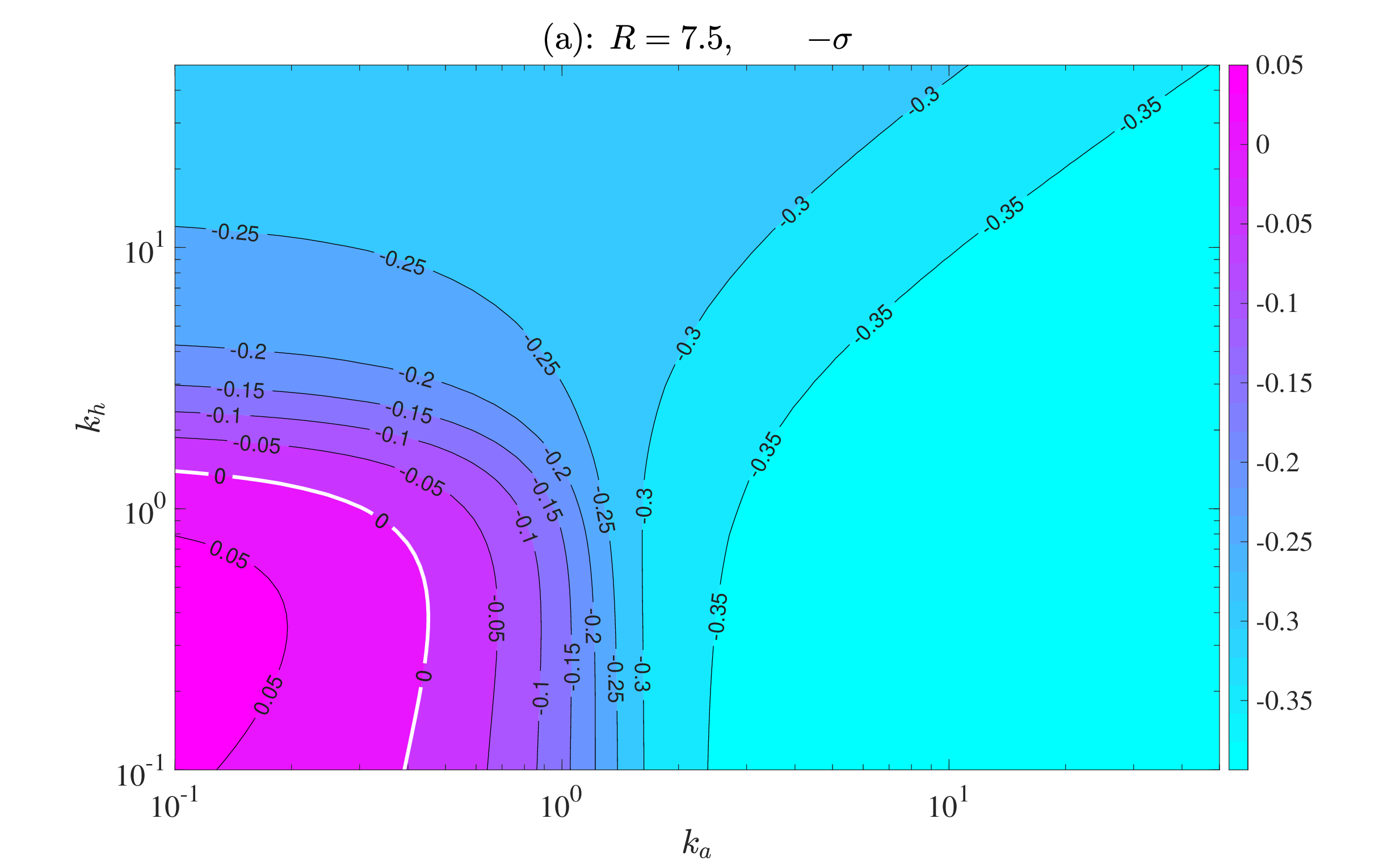,width=.55\linewidth}\epsfig{file=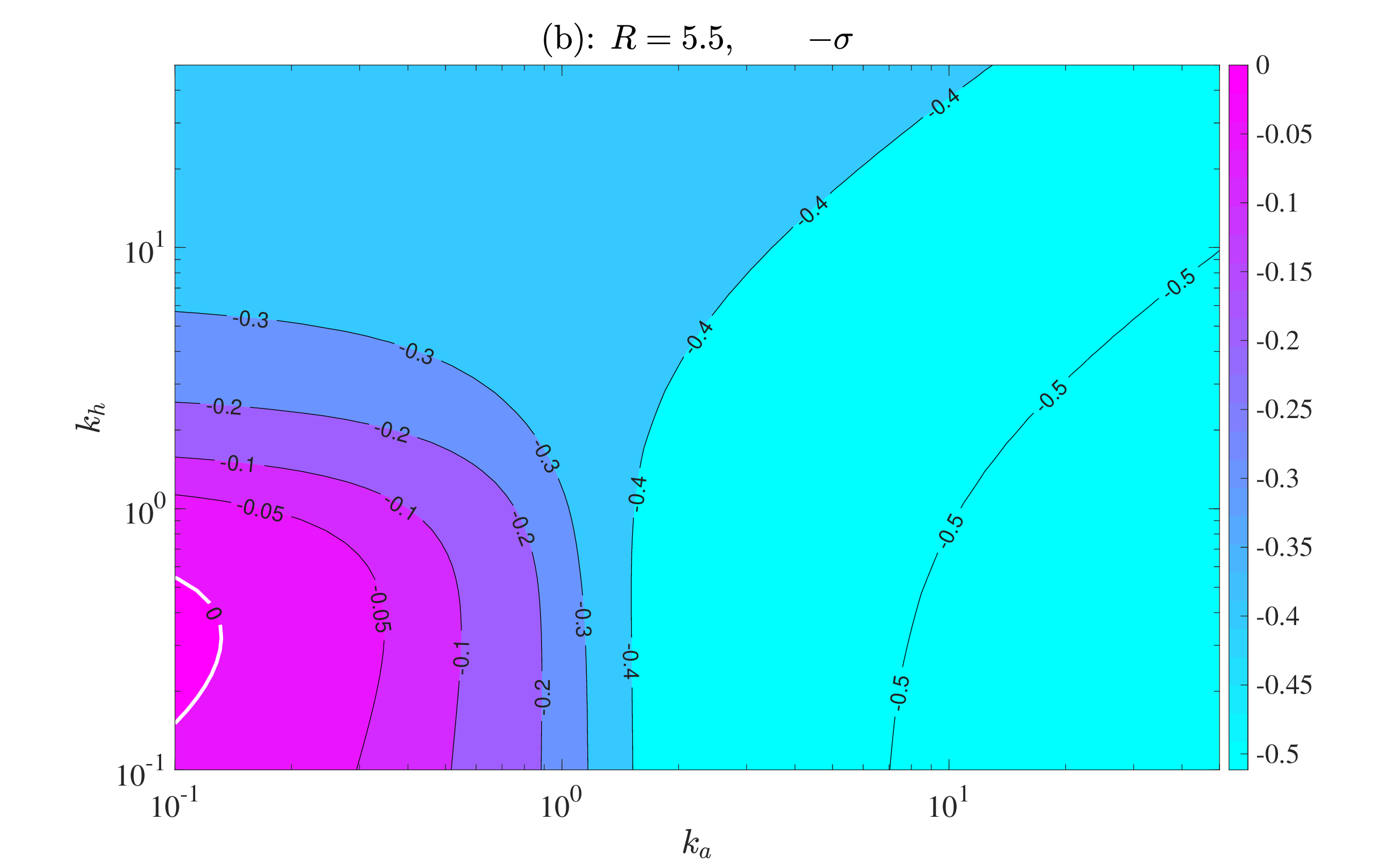,width=.55\linewidth}}
  \caption{As in Fig. \ref{Fig_res7}(b), but for $R=7.5$ (a) and $R=5.5$ (b).}
\label{Fig_res8}
\end{figure}

The region of flutter instabilities of a rigid foil in the $(k_a,k_h)$-plane, depicted in Fig \ref{Fig_res7} for $R=10$, shrinks when the mass ratio  $R$ decreases, as shown in Fig. \ref{Fig_res8} for $R=7.5$ and $R=5.5$. When $R=0.5$ (not shown),  the instability region is reduced to very small values of $k_a$, lying out  of the range plotted in Figs. \ref{Fig_res7}  and \ref{Fig_res8}.

\begin{figure}
\centerline{\epsfig{file=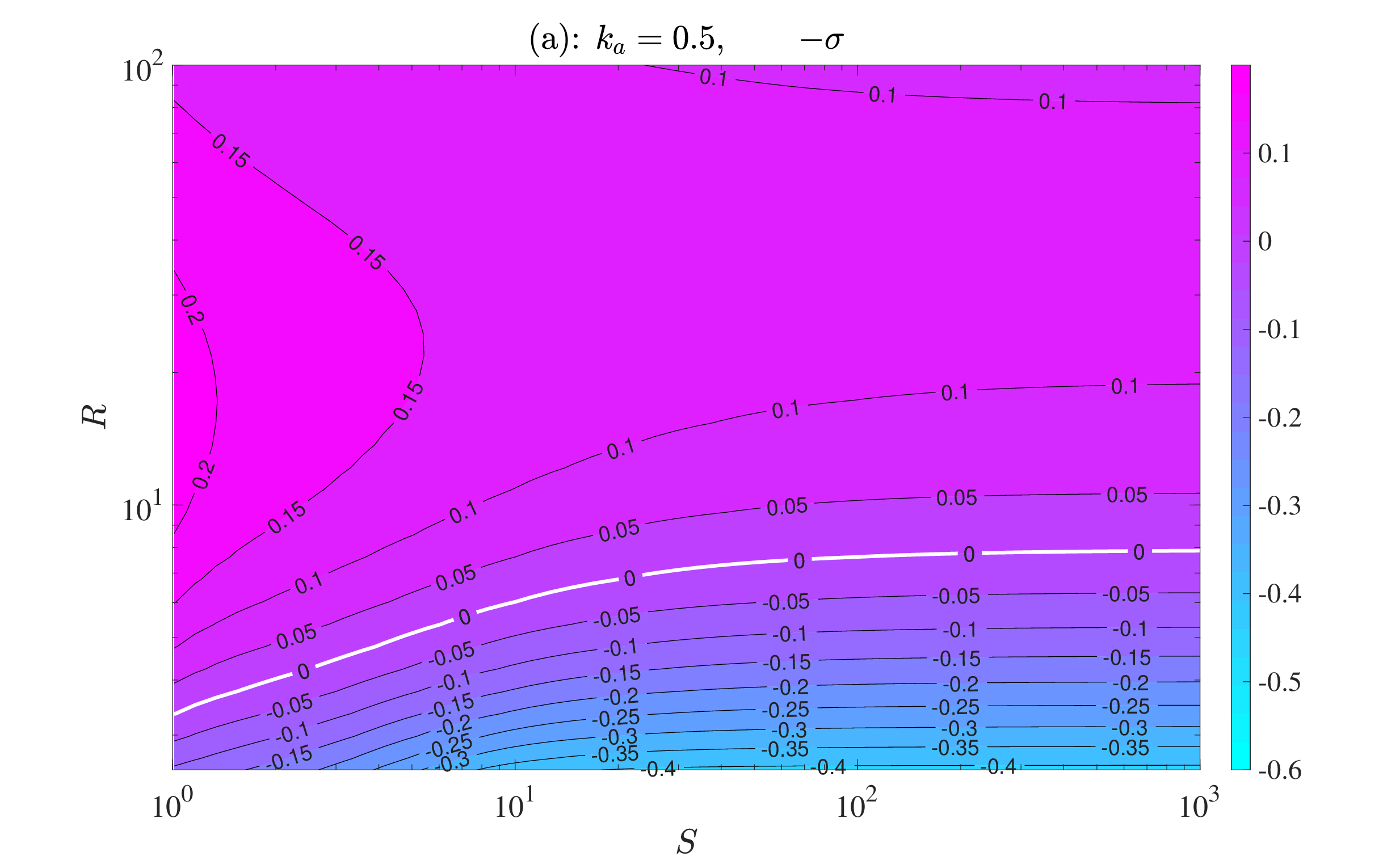,width=.55\linewidth}\epsfig{file=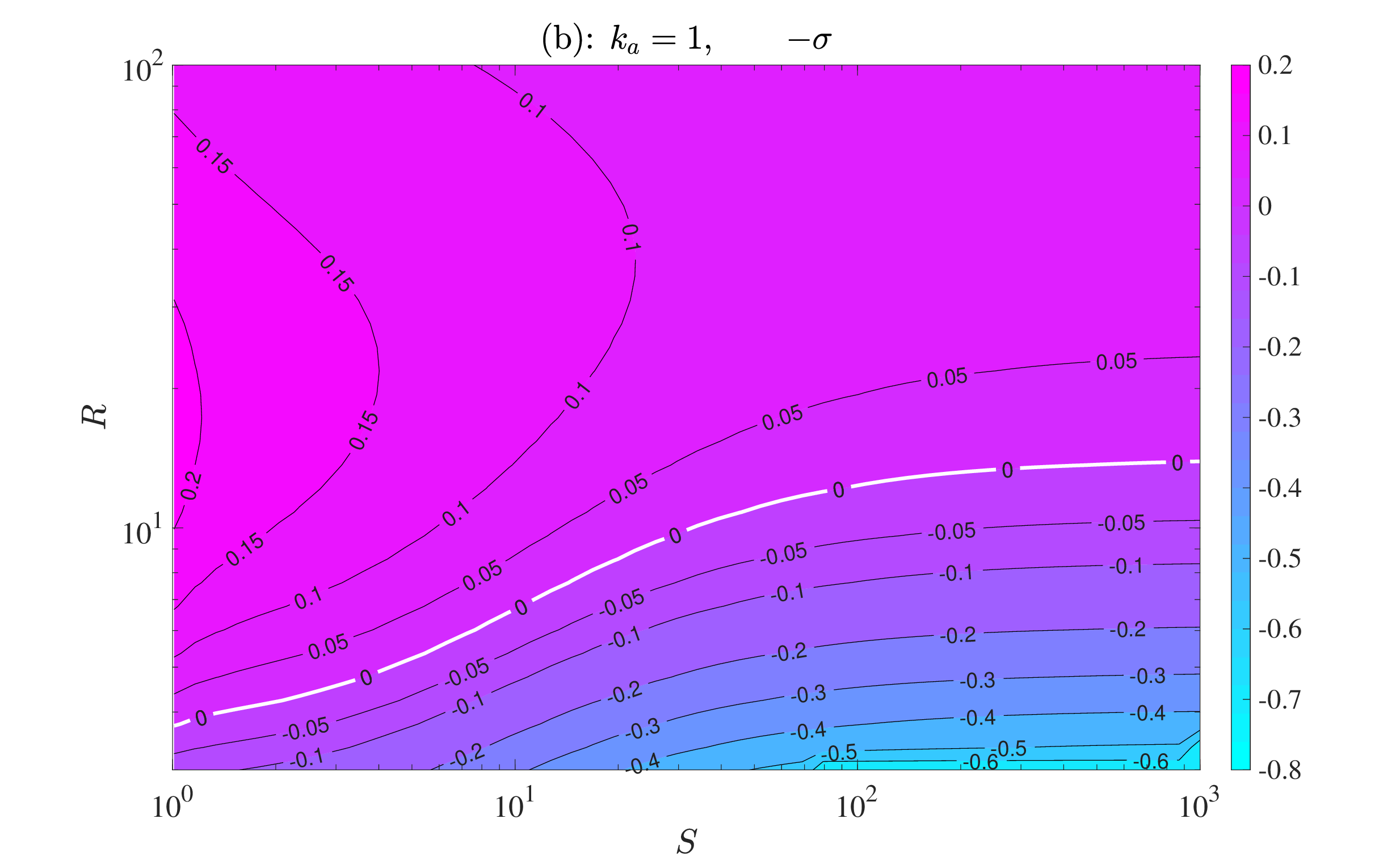,width=.55\linewidth}}
  \caption{Growth rate  for $k_h=0.4$  with $k_a=0.5$ (a) and $k_a$=1 (b) resulting from \eqref{detA0} when as $S$ and $R$ are varied.  The neutral curves $\sigma=0$ are marked with a thick white lines.}
\label{Fig_res9}
\end{figure}

To explore the effect of flexibility ($S$) and inertia ($R$) on these coupled pitch-heave unstable modes we select  $k_h=0.4$ and two values of $k_a$. This is made in Fig. \ref{Fig_res9}, where the the growth rate $-\sigma$, resulting from \eqref{detA0} using the minimum of $(k_r,k_f)$ as seed, are plotted on the $(R,S)$-plane for $k_a=0.5$ and $k_a=1$. The main effect of flexibility is to increase  the flutter  instability region by  lowering the minimum threshold value  of the mass ratio $R$  to support instabilities.  This effect is more accused as $k_a$ increases. As discussed above,  the threshold value of $R$ for coupled pitch-heave flutter instability of a rigid foil increases with $k_a$, decreasing the unstable region. This effect   can be observed  in Fig. \ref{Fig_res9} by comparing  the two cases plotted at, for instance,  $S=10^3$. As $S$ decreases, the spring mode of flutter instability becomes more and more coupled with the flexural instability mode (which is noticeable in the contour lines of the figure when they start to deflect when  $S$ decreases), enlarging the range of the mass ratio for flutter instability and increasing the growth rate. This effect of flexibility starts at higher values of $S$ as $k_a$ increases,  also expanding the instability region more quickly for larger values of $k_a$. As it is better appreciated in Fig. \ref{Fig_res10}(a), this enlargement of the instability region as $k_a$ increases is actually due to the rapid  increase of the threshold value for instability of a rigid foil, while the threshold value for small stiffness ($S=1$, say) remains almost constant.

\begin{figure}
\centerline{\epsfig{file=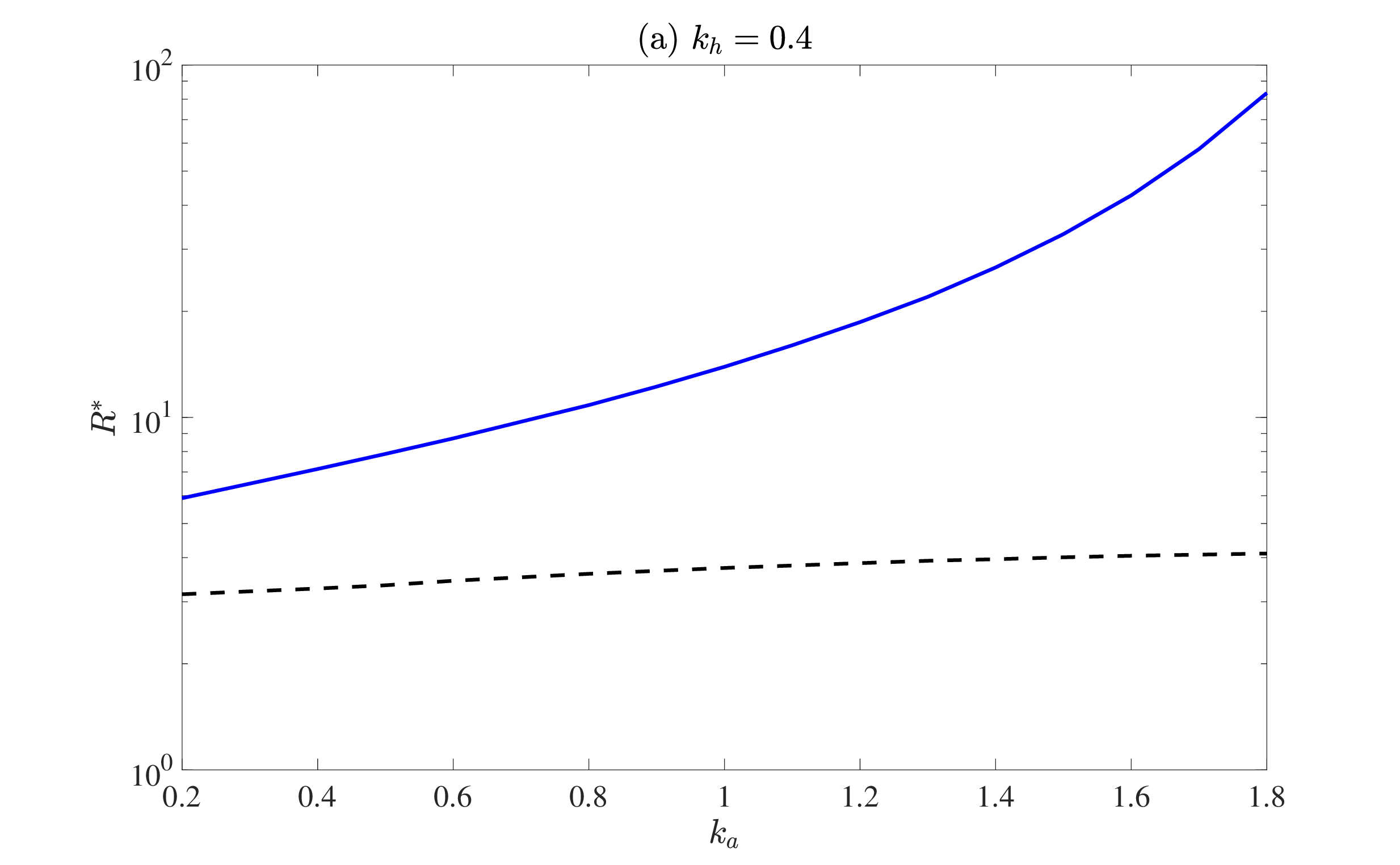,width=.55\linewidth}\epsfig{file=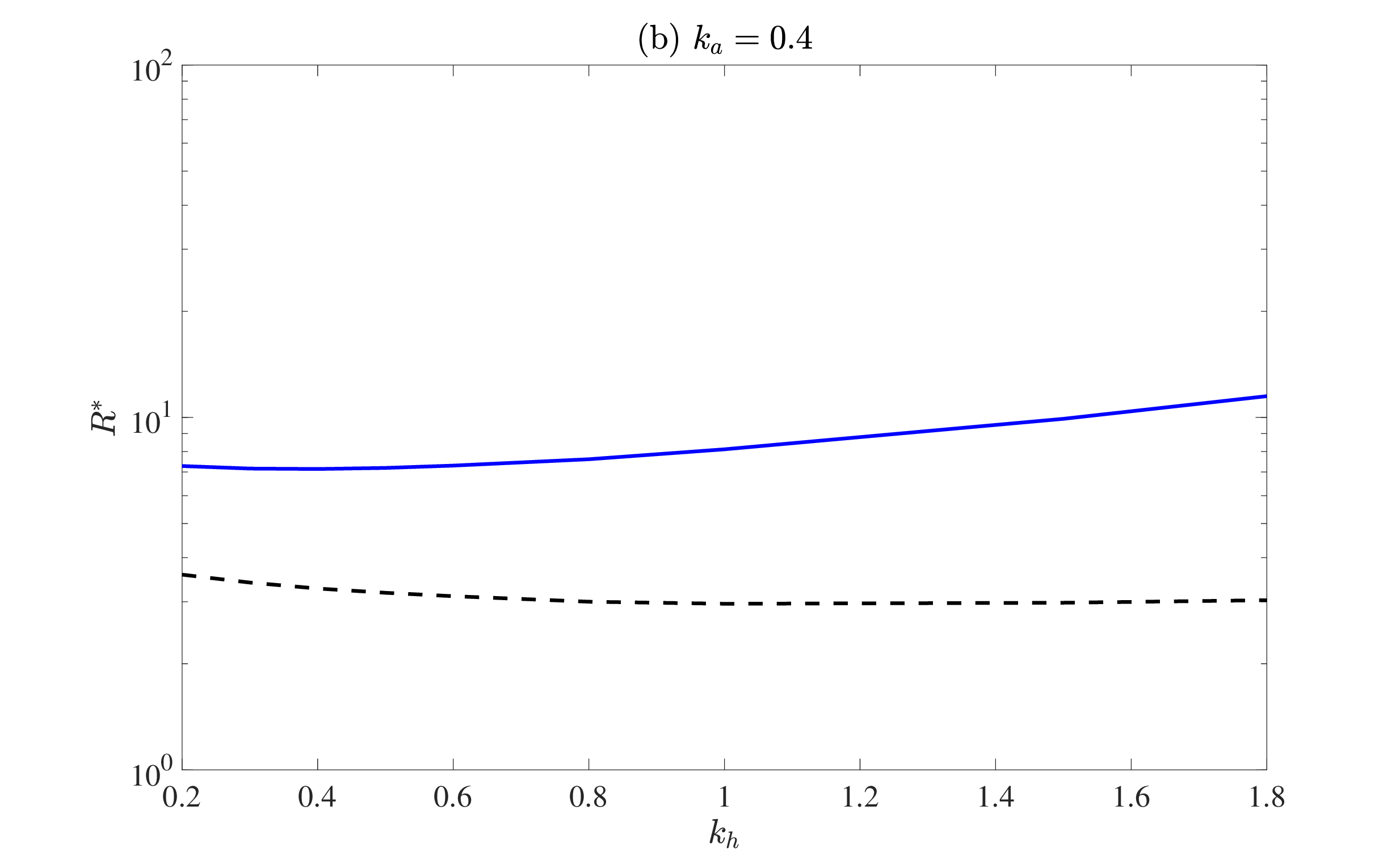,width=.55\linewidth}}
  \caption{Minimum mass ratio $R^*$ for instability ($\sigma=0$) for $S=1$ (continuous lines) and $S=10^3$ (dashed lines) when $k_h=0.4$ as $k_a$ is increased (a), and when $k_a=0.4$ as $k_h$ increases.}
\label{Fig_res10}
\end{figure}

Similar behavior occurs for a fixed value of $k_a$ as $k_h$ increases. But now the enlargement of the instability region as $k_h$ grows is much smaller because the  threshold of the mass ratio  for instability of the rigid foil increases much more slowly, as shown in Fig.  \ref{Fig_res10}(b) for $k_a=0.4$. However,  the mass ratio threshold for small stiffness ($S=1$) now decreases slightly when $k_h$ increases, thus helping to enlarge the instability region as the stiffness decreases for fixed values of $k_a$ and $k_h$.

\subsection{Effect of the dampers} \label{sec_dampers}

As mentioned above, most of the  results reported so far are for fixed values of the dampers constants, $b_a=b_h=0.5$. Obviously, the effect of decreasing (increasing) the dampers constants is to increase (decrease) both the instability parametric regions and the corresponding growth rates.

\begin{figure}
\centerline{\epsfig{file=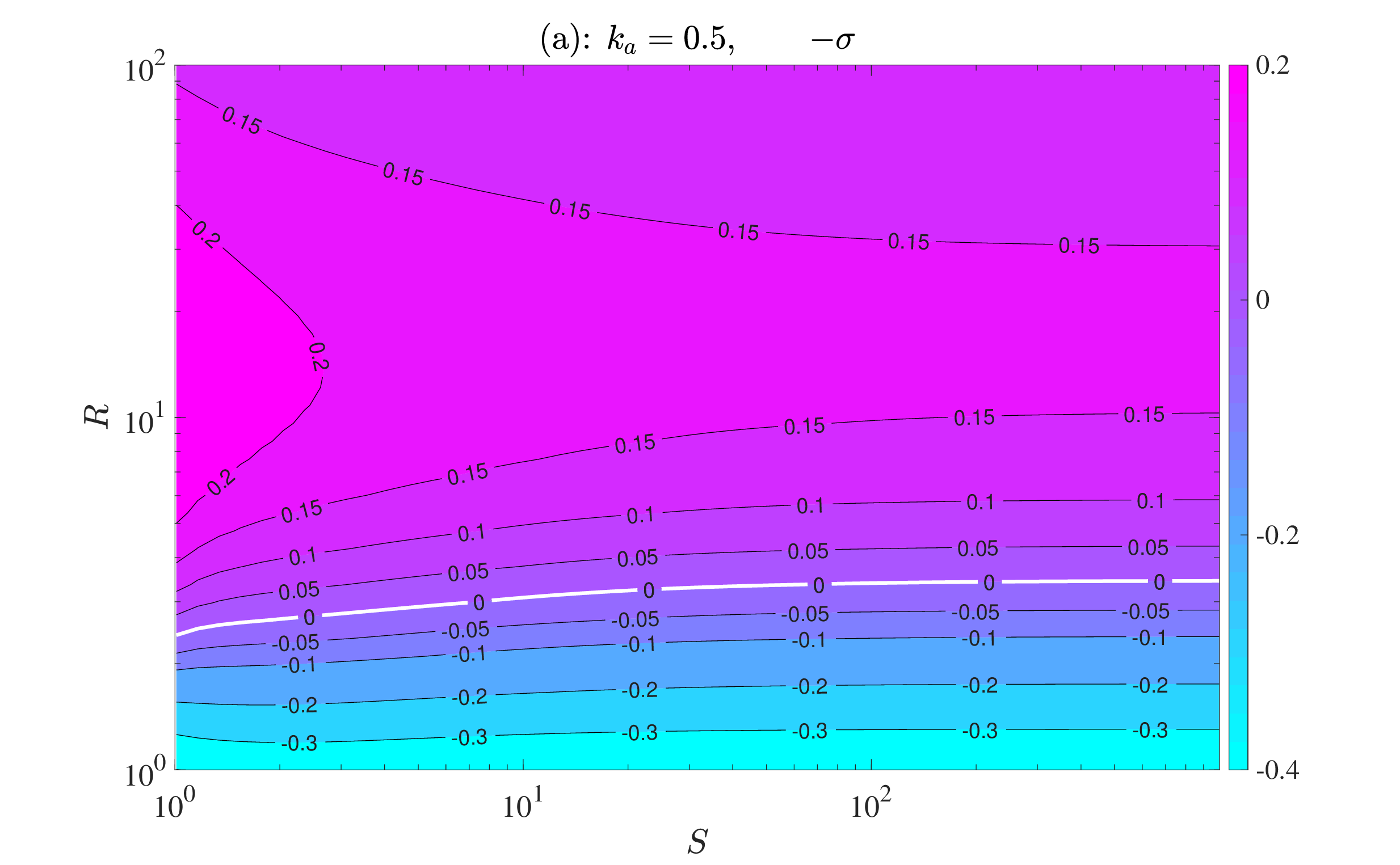,width=.55\linewidth}\epsfig{file=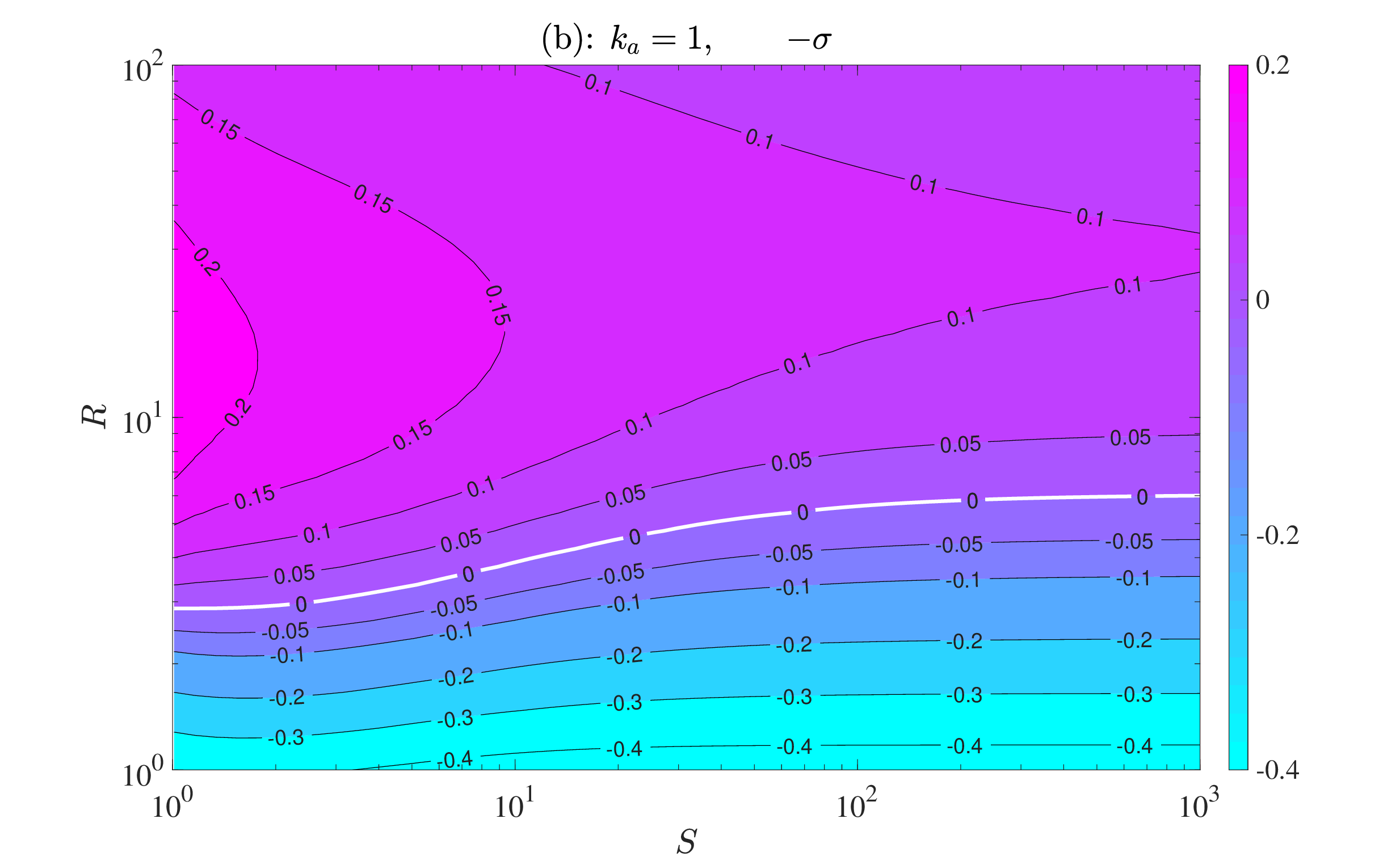,width=.55\linewidth}}
  \caption{As in Fig. \ref{Fig_res9}, but using $b_a=b_h=0$ instead of $b_a=b_h=0.5$. Note that a different scale has been used in the ordinate.}
\label{Fig_res11}
\end{figure}

To show these two effects with just one example, Fig. \ref{Fig_res11} shows the results for the same parameters used in Fig. \ref{Fig_res9}, but using $b_a=b_h=0$ instead of $b_a=b_h=0.5$. Now the foil is unstable at lower values of the mass ratio (note the different scale in the ordinate) and the growth rate for the same values of the remaining parameters is higher.

\section{Concluding remarks}
\label{sec_conclu}

The analytical tool presented here to predict the onset of  flutter instabilities in a flexible foil elastically mounted on springs and dampers at its leading edge has been validated with previous numerical results available for several particular configurations. The method captures accurately the first couple of flutter frequencies as the stiffness decreases from the rigid-foil limit, thus predicting the critical flutter velocities, and their corresponding frequencies,  down to values of the non-dimensional stiffness $S=4E (\varepsilon/c)^3/(\rho U^2)$ of order $10^{-1}$, but not the bunch of higher-frequencies  flapping-flag instabilities occurring at lower values of $S$. The analytical method is thus useful as a guide for the design of turbines based on oscillating flexible foils, among other applications where it is important to characterize the aeroelastic instabilities of not too flexible foils, taking into account  the coupling between pitch, heave and flexibility of the foil. The effect of gravity is also taken into account in the analysis. Although it does not affect directly  to the flutter stability in the present linearized formulation, the analysis provides a simple expression for  the equilibrium location and shape of the elastically mounted flexible foil immersed in the fluid current, which depend on the non-dimensional stiffnesses of the springs and the foil ($k_h$, $k_a$ and $S$) and on a parameter which is a combination of the mass ratio and the Froude number, $2 \varepsilon g (\rho_s-\rho)/(\rho U^2)$. This equilibrium  information may be useful as an initial condition to trigger numerical stability analyses.

The main results on the parametric regions identified for the flutter instabilities, that result from the coupling of the springs with the flexural deformation of the foil, and the critical flutter conditions, can be summarized as follows:

\begin{itemize}

\item For a clamped foil at its leading edge (very large spring rigidities $k_h$ and $k_a$) flutter instabilities are generated only above a critical value of the mass ratio $R$ that depends on the stiffness $S$, for all values of $S$, but with significantly high growth rates only for $S \lesssim 10$.

\item When only passive heave, or only passive pitch, is allowed, no flutter instabilities are possible without foil flexibility effects (they are not generated with a rigid foil, as is well known). As the linear, or torsional, spring constant decreases, the flexural instability mode becomes coupled with the corresponding spring instability mode below a threshold value of the stiffness $S$ that depends on the mass ratio $R$,  increasing significantly the growth rate. For a given value of the spring stiffness ($k_h$ or $k_a$), these coupled spring-flexural instability modes are confined in a relatively small parametric region above a threshold value of $R$ and below a threshold value of $S$, a region that is easily obtained with the present analytical formulation within its range of validity.

\item Foil flexibility lowers the  mass ratio threshold above which the coupled pitch-heave instability mode of a rigid foil is possible for given $k_h$ and $k_a$, specially for $S \lesssim 10$, where the strong coupling between the two springs and the flexural unstable modes also  increases substantially the growth rate.

\end{itemize}

\appendix

\numberwithin{equation}{section}

\section{Fluid force and moment coefficients}
\label{app_coeffi}
In the present linear potential flow approximation the fluid force and moment coefficients \eqref{defCL}-\eqref{defCF2}, for the general foil motion \eqref{zs_def} and harmonic functions 
\be  h(t) =  h_0 e^{i k t}  \,, \q \alpha(t) = \alpha_0 e^{i k t}  \,, \q  d_1(t) = d_{10} e^{i k t}  \,, \q  d_2(t) = d_{20} e^{i k t}  \,,\label{halphada} \ee
with reduced frequency
\be k = \frac{\omega  c}{2 U} \,, 
\ee
are given by \citep{ferna25}
\be C_{L}(t) =   \pi \left( \dot{\alpha} + \ddot{\alpha} - \ddot{h} - 25 \dt_1 - \frac{149}{8} \ddt_1 - \frac{1465}{8} \dt_2 - \frac{1073}{8} \ddt_2 \right) +  \Gamma_0(t) \C (k) \,, \label{CL0}
\ee
\[ C_{M}(t)= \frac{\pi}{256} \left( 192 \dot{\alpha} + 144 \ddot{\alpha} - 128 \ddot{h} - 576 d_1 -5248 \dt_1 - 2800 \ddt_1 - 4640 d_2   \right. \]
\be \q \q \q \q   \left. - 38680 \dt_2-20213 \ddt_2 \right) + \frac{1}{4} \Gamma_0 (t) \C(k) \,, \label{CM0}
\ee
\[ C_{F_{1}}(t)= \frac{\pi}{192} \left( 384 \dot{\alpha} + 288 \ddot{\alpha} - 240 \ddot{h} - 1056 d_1 -10848 \dt_1 - 5745  \ddt_1 - 8640 d_2 \right. \]
\be \q \q \q \q \left.  - 80190 \dt_2-41540 \ddt_2 \right) +  \frac{1}{2} \Gamma_0 (t) \C(k) \,, \label{CF10}
\ee
\[ C_{F_{2}}(t)= \frac{\pi}{128} \left( 368 \dot{\alpha} + 280 \ddot{\alpha} - 224 \ddot{h} - 912 d_1 -10580 \dt_1 - 5680 \ddt_1 - 7530 d_2 \right. \]
\be \q \q \q \q \left.  - 78350 \dt_2-41117 \ddt_2 \right) +  \frac{5}{8} \Gamma_0 (t) \C(k)  \,, \label{CF20}
\ee
with
\be \Gamma_0(t)= \pi \left[ - 2 \dot{h} + 3 \dot{\alpha} + 2 \alpha - \frac{263}{4} \dt_1 - 59 d_1 - \frac{3831}{8} \dt_2 - \frac{1755}{4} d_2 \right] \,, \label{G0t}
\ee
and $\C(k)$ is Theodorsen's function. 
The buoyancy forces \eqref{CFe} are not included.

\section{Fluid force and moment coefficients for a stationary foil and equilibrium foil parameters}
\label{app_coeffi_e}
For a stationary foil, the fluid force and moments can be derived from the expressions in Appendix \ref{app_coeffi} by setting all the time derivatives to zero and taking into account that  $\C(k =0)=1$: 
\be C^*_{L_e} =  \pi \left( 2 \alpha - 59 d_1 - \frac{1755}{4} d_2 \right) \,, \label{clea}
\ee
\be C^*_{M_e}= - \frac{\pi}{8} \left( 18 d_1 + 145 d_2  \right) + \frac{\pi}{4} \left( 2 \alpha - 59 d_1 - \frac{1755}{4} d_2 \right) \,, 
\ee
\be C^*_{F_{1e}}= -  \frac{\pi}{2} \left(  11 d_1  + 90 d_2 \right) + \frac{\pi}{2} \left( 2 \alpha - 59 d_1 - \frac{1755}{4} d_2 \right) \,,
\ee
\be C^*_{F_{2e}}= - \frac{\pi}{64} \left(  456 d_1 + 3765 d_2 \right) + \frac{5 \pi}{8} \left( 2 \alpha - 59 d_1 - \frac{1755}{4} d_2 \right) \,. \label{cfea}
\ee
Note that the last terms between large brackets come from the circulatory contributions, i.e. from the terms containing $\Gamma_0$ in Appendix \ref{app_coeffi}, while the remaining ones in $C^*_{M_e}$, $C^*_{F_{1e}}$ and $C^*_{F_{2_e}}$ come from the non-circulatory terms.

Substituting these expressions into Eqs.  \eqref{mom1a}-\eqref{mom4a}, together with \eqref{CFe}  and \eqref{CLpCMp}, one obtains a set of four linear equations whose solution yields the following  equilibrium values, which are more precise than \eqref{hea}-\eqref{d1ea} because they take into account the (small) effect of the non-circulatory moments:

\be h_e^*= \left( \frac{G}{k_h} \right) \frac{-16384  k_a S^2+66592 \pi  k_a S-9255 \pi ^2  k_a+8192 \pi   S^2-2416 \pi ^2  S+75 \pi ^3 }{16384 k_a S^2+54240 \pi  k_a S+4635 \pi ^2 k_a+8192\pi  S^2+1168 \pi ^2 S+75 \pi ^3/2} \,, \label{heaa}
\ee
\be \alpha_e^* = \left(\frac{G}{2}\right)  \frac{16384 S^2-15392 \pi  S+1235 \pi ^2}{16384 k_a S^2+54240 \pi  k_a S+4635 \pi ^2 k_a+8192\pi  S^2+1168 \pi ^2 S+75 \pi ^3/2},
\ee
\be d_{1e}^* = \left(\frac{G}{2}\right) \frac{-4096  k_a S+32520 \pi   k_a+3328 \pi   S+395 \pi ^2 }{16384 k_a S^2+54240 \pi  k_a S+4635 \pi ^2 k_a+8192\pi  S^2+1168 \pi ^2 S+75 \pi ^3/2},
\ee
\be d_{2e}^* = - \left(\frac{G}{2}\right)  \frac{16 \pi  (272  k_a+32 G S+3 \pi  )}{16384 k_a S^2+54240 \pi  k_a S+4635 \pi ^2 k_a+8192\pi  S^2+1168 \pi ^2 S+75 \pi ^3/2} \,. \label{d2eaa}
\ee

\section{Terms in matrix $\tA$ in  \eqref{linearA}}
\label{app_coeffA}

\be A_{hh}= -R \gamma^2 +k_h + b_h i\gamma + \pi \left[ -\gamma^2 +2 i \gamma \C(\gamma) \right]  \,, \label{ahh} 
\ee
\be A_{ha}= R \gamma^2 +  \pi \left[-  i\gamma +\gamma^2 -\C(\gamma) \left(3 i\gamma + 2 \right) \right]  \,, 
\ee
\be A_{h1}= - \frac{96}{5} R \gamma^2  +  \pi \left[ 25 i \gamma  - \frac{149}{8} \gamma^2   + \C(\gamma) \left( \frac{263}{4}  i \gamma + 59 \right)  \right]  \,, 
\ee
\be A_{h2}= - \frac{416}{3} R \gamma^2  +  \pi \left[ \frac{1465}{8} i \gamma  - \frac{1073}{8} \gamma^2   + \C(\gamma) \left( \frac{3831}{8}  i \gamma + \frac{1755}{4} \right)\right]  \,, 
\ee
%************
\be A_{ah}= - \frac{1}{2} R \gamma^2 + \pi \left[ - \frac{1}{2} \gamma^2 + \frac{1}{2} i \gamma \C(\gamma) \right]  \,,
\ee
\be A_{aa}=  \frac{2}{3} R \gamma^2 - k_a - b_a i\gamma  +  \pi \left[- \frac{3}{4} i\gamma + \frac{9}{16} \gamma^2 - \frac{1}{4} \C(\gamma) \left(3 i\gamma + 2 \right) \right]  \,, 
\ee
\be A_{a1}= - \frac{208}{15} R \gamma^2  +  \pi \left[ \frac{9}{4} + \frac{1321}{64} i \gamma  - \frac{175}{16} \gamma^2   + \frac{1}{4} \C(\gamma) \left( \frac{263}{4}  i \gamma + 59 \right)  \right]  \,, 
\ee
\be A_{a2}= - \frac{704}{7} R \gamma^2  +  \pi \left[ \frac{145}{8} +\frac{4835}{32} i \gamma  - \frac{20213}{256} \gamma^2   + \frac{1}{4} \C(\gamma) \left( \frac{3831}{8}  i \gamma + \frac{1755}{4} \right)\right]  \,, 
\ee
%************
\be A_{1h}= - \frac{4}{3} R \gamma^2 + \pi \left[ - \frac{5}{4} \gamma^2 + i \gamma \C(\gamma) \right]  \,,
\ee
\be A_{1a}=  2 R \gamma^2 +  \pi \left[- 2 i\gamma + \frac{3}{2} \gamma^2 - \frac{1}{2} \C(\gamma) \left(3 i\gamma + 2 \right) \right]  \,, 
\ee
\be A_{11}= - \frac{4544}{105} R \gamma^2  + \frac{32}{3} S +  \pi \left[ \frac{11}{2} + \frac{113}{2} i \gamma  - \frac{5745}{192} \gamma^2   + \frac{1}{2} \C(\gamma) \left( \frac{263}{4}  i \gamma + 59 \right)  \right]  \,, 
\ee
\be A_{12}= - \frac{944}{3} R \gamma^2   + 80 S +  \pi \left[ 45 +\frac{13365}{32} i \gamma  - \frac{10385}{48} \gamma^2   + \frac{1}{2} \C(\gamma) \left( \frac{3831}{8}  i \gamma + \frac{1755}{4} \right)\right]  \,, 
\ee
%************
\be A_{2h}= - 2 R \gamma^2 + \pi \left[ - \frac{7}{4} \gamma^2 + \frac{5}{4}  i \gamma \C(\gamma) \right]  \,,
\ee
\be A_{2a}=  \frac{16}{5}  R \gamma^2 +  \pi \left[- \frac{23}{8} i\gamma + \frac{35}{16} \gamma^2 - \frac{5}{8} \C(\gamma) \left(3 i\gamma + 2 \right) \right]  \,, 
\ee
\be A_{21}= - \frac{496}{7} R \gamma^2  + 16 S +  \pi \left[ \frac{57}{8} + \frac{2645}{32} i \gamma  - \frac{355}{8} \gamma^2   + \frac{5}{8} \C(\gamma) \left( \frac{263}{4}  i \gamma + 59 \right)  \right]  \,, 
\ee
\be A_{22}= - \frac{32512}{63} R \gamma^2   + 128 S +  \pi \left[ \frac{3765}{64} +\frac{39175}{64} i \gamma  - \frac{41117}{128} \gamma^2   + \frac{5}{8} \C(\gamma) \left( \frac{3831}{8}  i \gamma + \frac{1755}{4} \right)\right]  \,. \label{a22} 
\ee
Note that the fluid contributions are the terms in square brackets multiplied by $\pi$, with the circulatory contributions those factorized with Theodorsen's function $\C(\gamma)$. Inertia and stiffness contributions ara characterized by the non-dimensional parameters $R$ and $S$, respectively, whereas the springs and dampers terms are those containing the cosntants $k_h$, $b_h$, $k_a$ and $b_a$.  Except for the coefficients affecting the second flexural mode $d_2$, i.e. $A_{i2}$ and $A_{2,j}$, $i=h,a,1$ and $j=h,a,1,2$, which are missing in the analysis of \cite{ferna22}, the remaining ones coincide with those reported in that reference if one takes into account that here, following the notation in \cite{ferna25},  $R=m$, $d_1=d/24$, $a=-1$ and $x_0=0$. \\

\noindent {\bf{Declaration of competing interest}}\\

The author reports no conflict of interest.\\

\noindent {\bf{Acknowledgments}}\\

This research has been supported by the MCIyU/AEI grant PID2023-150588NB-I00.\\

 \bibliographystyle{model1-num-names}
%\bibliographystyle{cas-model2-names}

%\bibliography{referen}


\end{document}